\documentclass[lettersize,journal]{IEEEtran}
\usepackage{amsmath,amsfonts}
\usepackage{algorithmic}
\usepackage{algorithm}
\usepackage{array}
\usepackage[caption=false,font=normalsize,labelfont=sf,textfont=sf]{subfig}
\usepackage{textcomp}
\usepackage{stfloats}
\usepackage{url}
\usepackage{verbatim}
\usepackage{graphicx}
\usepackage{cite}
\begin{document}

\title{Efficient and Privacy-Preserving Federated Learning based on Full Homomorphic Encryption}
\author{Yuqi Guo,~\IEEEmembership{Member,~IEEE, } Lin Li,~\IEEEmembership{Member,~IEEE, } Zhongxiang Zheng,~\IEEEmembership{Member,~IEEE, }Hanrui Yun,~\IEEEmembership{Member,~IEEE, }  Ruoyan Zhang, ~\IEEEmembership{Member,~IEEE, } Xiaolin Chang,~\IEEEmembership{Senior Member,~IEEE, }Zhixuan Gao, ~\IEEEmembership{Member,~IEEE }
        % <-this % stops a space
\thanks{
Yuqi Guo, Lin Li, Xiaolin Chang are with the school of Computer Information and Technology, Beijing Jiaotong University, Beijing 100044, China. E-mail: 22120492@bjtu.edu.cn, lilin@bjtu.edu.cn, xlchang@bjtu.edu.cn. Lin Li is the corresponding author.
}
\thanks{
Zhongxiang Zheng, Hanrui Yun and Ruoyan Zhang are with the school of Computer and Cyber Science, Communication University of China, Beijing 100024, China. E-mail: zhengzx@cuc.edu.en, hanruiyun@cuc.edu.cn, 1742223174@qq.com. Zhongxiang Zheng is also the corresponding author.
}
\thanks{
Zhixuan Gao is with the school of Cyberspace Security, Beijing Institute of Technology, Beijing 100081, China. E-mail: gzx20000617@gmail.com.
}
}% <-this % stops a space
%\thanks{Manuscript received April 19, 2021; revised August 16, 2021.}}



\maketitle

\begin{abstract}
Since the first theoretically feasible full homomorphic encryption (FHE) scheme was proposed in 2009, great progress has been achieved. These improvements have made FHE schemes come off the paper and become quite useful in solving some practical problems.  In this paper, we propose a set of novel Federated Learning Schemes by utilizing the latest homomorphic encryption technologies, so as to improve the security, functionality and practicality at the same time. 
\par Comparisons have been given in four practical data sets separately from medical, business, biometric and financial fields, covering both horizontal and vertical federated learning scenarios. The experiment results show that our scheme achieves significant improvements in security, efficiency and practicality, compared with classical horizontal and vertical federated learning schemes.
\end{abstract}


\section{Introduction}
\setlength{\parskip}{0.5em}
\IEEEPARstart{W}{ith} the ever-expanding collection and use of data, data already has a high economic value and has been called the oil of the digital economy. When bringing economic benefits, data also brings the hidden danger of personal privacy leakage at the same time. As a result, the research of privacy-preserving computing has attracted more and more attentions from  researchers in both  academia and industry. Privacy-preserving computing is defined as the techniques can be used to provide a balance of data usage and privacy protection, which include federated learning \cite{DBLP:conf/aistats/McMahanMRHA17}, secure multi-party computation\cite{MEMBER1981On}, trusted execution environment \cite{10.1145/1866307.1866390}, differential privacy\cite{dwork2006calibrating}, and homomorphic encryption \cite{RonaldRivest}. Among them, federated learning and homomorphic encryption are two key technologies for privacy-preserving computing.
\par We propose a set of federated learning schemes based on full homomorphic encryption (FHE). The core idea of our schemes is to comprehensively improve the classical federated learning schemes by using the latest homomorphic encryption technologies, so as to improve the security, efficiency and practicality.
\par In classical federated learning schemes, a partially homomomorphic encryption (PHE) algorithm named Paillier acts as an important basic module. We show that by replacing PHE algorithm by FHE algorithm and carefully designing the schemes, significant improvements can be achieved in federated learning. The reasons can be summarized as three aspects.
\par Firstly, when encrypted by FHE, the ciphertexts can conduct additions or multiplications directly which is the main feature  provided by FHE. However, when using PHE, either ciphertext addition or ciphertext multiplication is available. For example, the Paillier algorithm supports ciphertext addition and scalar multiplication, and does not support ciphertext multiplication. While a  FHE algorithm  supports both ciphertext multiplication and ciphertext addition at the same time. Using this property, we can redesign the federated learning schemes in a different way, e.g., a unified horizontal/vertical federated learning framework. 
\par Secondly, security is another key issue. Currently, quantum computing has posed a fatal threat to a large number of classical algorithms, such as RSA, ECDSA, Paillier, etc.  and the anti-quantum performance of FHE algorithms is the key to ensure the continued availability of privacy-preserving computing solutions in the post-quantum era. Concurrently, there is a rise in gradient attacks and side-channel attacks targeting classical models, enabling attackers to access sensitive data via the gradient and information generated during model execution. The FHE algorithm can provide controllable security for models and data, which is an important method for privacy protection.
\par Thirdly, in the recent years, with the rapid developments of new technologies, the efficiency of FHE algorithms has been improved significantly, for example,  our scheme has achieved a great   efficiency improvement in training modules compared with classical schemes using PHE algorithms, which is a strong evidence that FHE algorithms have sufficient practicality.
\par Using the above technical features of FHE, we redesign the SecureBoost model\cite{DBLP:journals/expert/ChengFJLCPY21} and the Logistic Regression (LR) model\cite{olr} for horizontal and vertical federated learning, and propose a new set of federated learning schemes based on FHE, which achieve the following advantages over the classical federated learning schemes.

\par \textbf{More Functions:} Our federated learning schemes support both tree model and linear model. At the same time, based on the functionality of the FHE algorithm, our federated learning schemes can support high precision approximation of complex loss functions and can cope with more complex training objectives and tasks. Our federated learning schemes provide protection of models and data for a wider range of application scenarios, specific functions and advantages are shown in Table \ref{advantage}. Our SecureBoost model for vertical federated learning provides a more secure and less communicative approach to federated inference, and our Logistic Regression model for vertical federated learning can support model evaluation operations initiated by any participant. In addition, our proposed federated learning scheme based on FHE achieves unification in the training process, and participants can perform horizontal/vertical federated learning operations according to data distribution in the same framework. Therefore, there is no need to deploy separate horizontal/vertical versions. We also implement the computation of WOE values and the SMOTE algorithm based on FHE, allowing our federated learning to be applied in scenarios where the dataset is highly unbalanced between positive and negative samples, which is not possible with classical federated learning. 

\begin{table*}[]
\centering
\caption{Functions achieved by our federated learning schemes}
\label{advantage}
\begin{tabular}{c|c|c|c|c|c|c|c|c|c}
\hline
Model                                                  & \begin{tabular}[c]{@{}c@{}}High\\ efficiency\end{tabular} & \begin{tabular}[c]{@{}c@{}}Unified horizontal\\ /vertical versions\end{tabular} & \begin{tabular}[c]{@{}c@{}}Federated\\ SMOTE\end{tabular} & \begin{tabular}[c]{@{}c@{}}Model\\ evaluation\\ security\end{tabular} & \begin{tabular}[c]{@{}c@{}}Anti-gradient\\ attack\\ security\end{tabular} & Model                                                           & \begin{tabular}[c]{@{}c@{}}High\\ efficiency\end{tabular} & \begin{tabular}[c]{@{}c@{}}Low\\ communication\\ overheads\end{tabular} & \begin{tabular}[c]{@{}c@{}}Resistant to\\ side channel\\ attacks\end{tabular} \\ \hline
\begin{tabular}[c]{@{}c@{}}Classical\\ federated LR\end{tabular} & $\circ$                                                     & $\circ$                                                                           & $\circ$                                                     & $\circ$                                                                 & $\circ$                                                                     & \begin{tabular}[c]{@{}c@{}}Classical\\ SecureBoost\end{tabular} & $\circ$                                                     & $\circ$                                                                   & $\circ$                                                                         \\ \hline
Our LR                                                 & $\bullet$                                                      & $\bullet$                                                                            & $\bullet$                                                      & $\bullet$                                                                  & $\bullet$                                                                      & \begin{tabular}[c]{@{}c@{}}Our\\ SecureBoost\end{tabular}       & $\bullet$                                                      & $\bullet$                                                                    & $\bullet$                                                                          \\ \hline
\end{tabular}
\end{table*}

\par \textbf{Better Security:} The security of our federated learning schemes rely on the lattice problem, which is considered to have good resistant against quantum computing attacks and shares the same security assumption adopted by the post-quantum public key standard published by the National Institute of Standards and Technology (NIST), so that our schemes can provide protection in future quantum computing environments. In addition, CKKS algorithm provides controlled error protection for the data, so our federated learning schemes are as effective as the classical federated learning schemes with differential privacy protections in terms of security against gradient attacks, and can better protect the data of each participant.
\par \textbf{More efficiency:} With a well-designed algorithmic process and precise parameter selection, our proposed federated learning schemes obtain significant improvements in training efficiency over the classical federated learning schemes. Experimental results show that our secureboost federated learning model is 1.4-2 times more efficient than the classical algorithm, the horizontal logistic regression federated learning model is 9.3 times more efficient than the classical algorithm for training, and the vertical logistic regression federated learning model is 3-3.7 times more efficient than the classical algorithm for training.

\section{Background}
\setlength{\parskip}{0.5em}
\subsection{Notations}
\par First, we give notations used in this paper.
\begin{itemize}
\item Vectors are represented in bold lowercase letters (e.g.: $\mathbf{a}, \mathbf{b}, \mathbf{v}$),   matrixes are represented in bold capital letters, (e.g. : $\mathbf{A}, \mathbf{B}$). For the given set$~D$, The entire $~m$-dimensional vector whose elements belong to the set $~D$ is denoted $~D^m$, The elements belong to the entire $~m\times n$-order matrix of the set $~D$ is denoted $~D^{m\times n}$. 
\item For an $~n$-dimensional vector $~\mathbf{v}$, its $~i$-th element is denoted as $~\mathbf{v}_i~(0\leq i<n)$. 

\item  For an $~m\times n$   matrix$~\mathbf{A}$, its ($i,j$-th) element (i.e., the element in column $~j$ of row $~i$) is denoted as$~\mathbf{A}_{i,j}$($0\leq i<m~$ and $~0\leq j<n$), The $~i$-th row of the matrix is represented as $~\mathbf{A}_i=(\mathbf{A}_{i,0},  \mathbf{A}_{i,1}, \cdots,\mathbf{A}_{i,n-1})$. 

\item For a probability distribution $~\chi$, the symbol $~e\leftarrow \chi$ denotes that $~e$ is obtained by sampling according to the distribution $~\chi$. $\chi^n$  denotes $~n$ mutually independent $~\chi$. A normal distribution is denoted by $N(\mu,\sigma^2)$, where $\sigma$ is the standard deviation and $\mu$ represents its mean value.

\item For a real vector $~\mathbf{v}\in \mathbb{R}^n$, its Euclidean norm is denoted by $~\|\mathbf{v}\|$.

\item For two $~n$-dimensional vectors $~\mathbf{a},\mathbf{b}$ on the ring $~R$, the inner product is denoted as $~\langle \mathbf{a}, \mathbf{b}\rangle=\sum_{i=0}^{n-1}a_i\cdot b_i\in R$.

\item The privacy budget in differential privacy is denoted by $\epsilon$. The larger the privacy budget, the better the accuracy of the model but the larger the amount of privacy leakage.

\item Denote a negligible function by $negl\left(\lambda\right)$, i.e., one that satisfies $f\left(\lambda\right <1/p\left(\lambda\right)$ for any polynomial $p\left(\cdot\right)$ and a sufficiently large $\lambda$.
\end{itemize}

\subsection{Federated Learning}
 
The term federated learning (FL) was introduced in 2016 by McMahan et al \cite{DBLP:conf/aistats/McMahanMRHA17}, it involves multiple subject areas such as machine learning, artificial intelligence, and cryptography. It allows multiple clients to jointly train a model under the coordination of a central server, and the training data can be stored locally in a decentralized manner\cite{2021Advances}. According to the different overlapping relationships between data feature space and sample space of different data owners, the federated learning can be classified into Horizontal Federated Learning (HFL), Vertical Federated Learning (VFL), and Federated Transfer Learning (FTL)\cite{Li_2023, XIA2021100008}. In this paper, we focus on horizontal and vertical Federated Learning which are briefly described below.
\par
\begin{figure}[!h]
\centering
\includegraphics[width=3.4in]{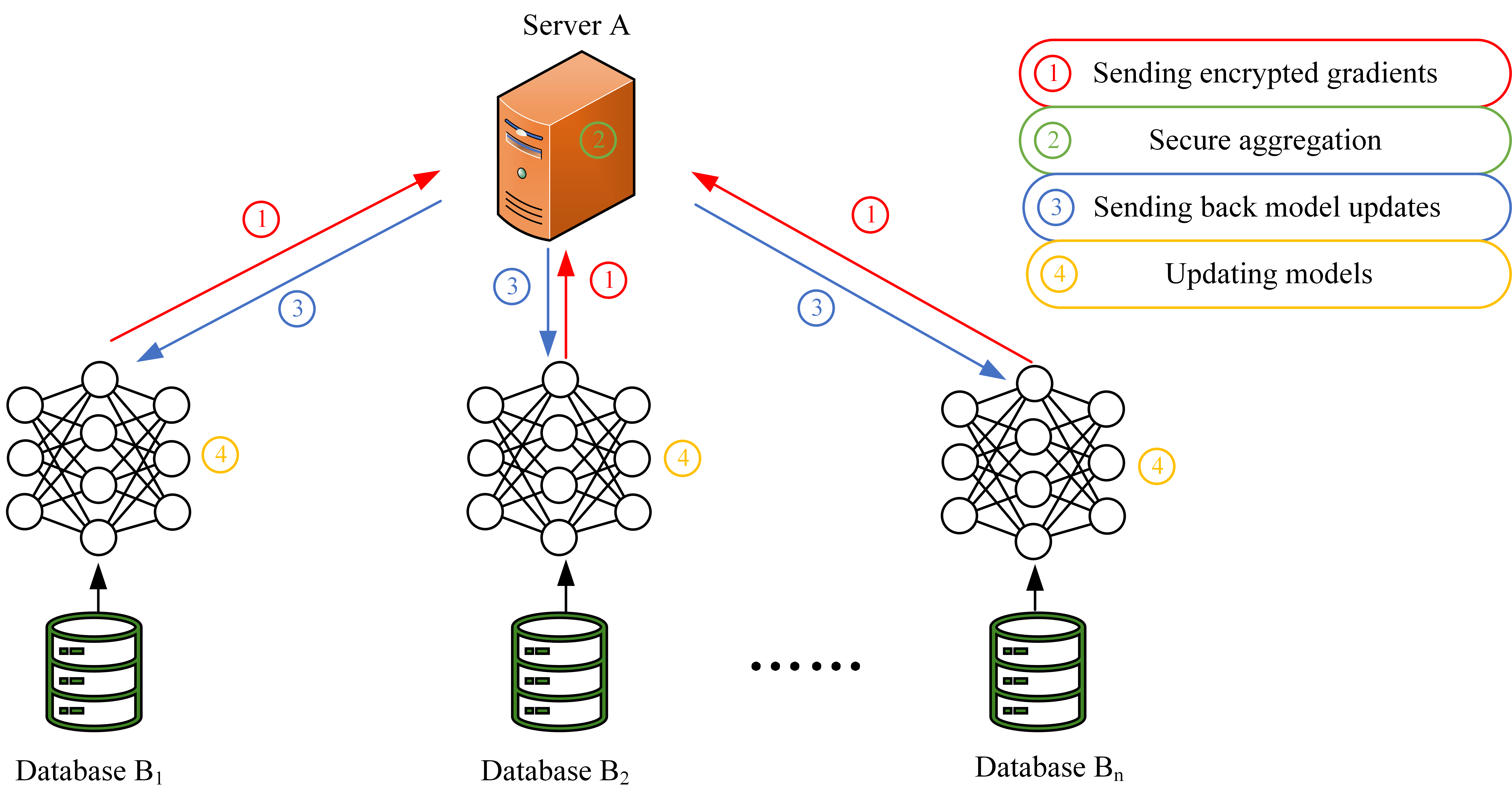}
\caption{Horizontal federated learning framework.}
\label{fig_4}
\end{figure}

 HFL is applicable to the case where the data features of the federated learning participants overlap more, i.e., the data features are aligned among the participants, but the data samples owned by the participants are different. Therefore, HFL is also known as sample-divided federated learning\cite{DBLP:journals/ftml/KairouzMABBBBCC21}, its framework is shown in Figure \ref{fig_4}. 
\par HFL was first proposed and is widely used in finance, medicine, IT, aviation, and other fields. Feng et al.\cite{DBLP:journals/tii/FengLYGW22} propose a block-chain empowered federated learning framework, and present its potential application scenarios in beyond 5G. Wang et al.\cite{DBLP:journals/tvcg/WangCXWZS23} develop a visual analytics tool for participating clients to explore data heterogeneity.
\par
\begin{figure}[!h]
\centering
\includegraphics[width=3.4in]{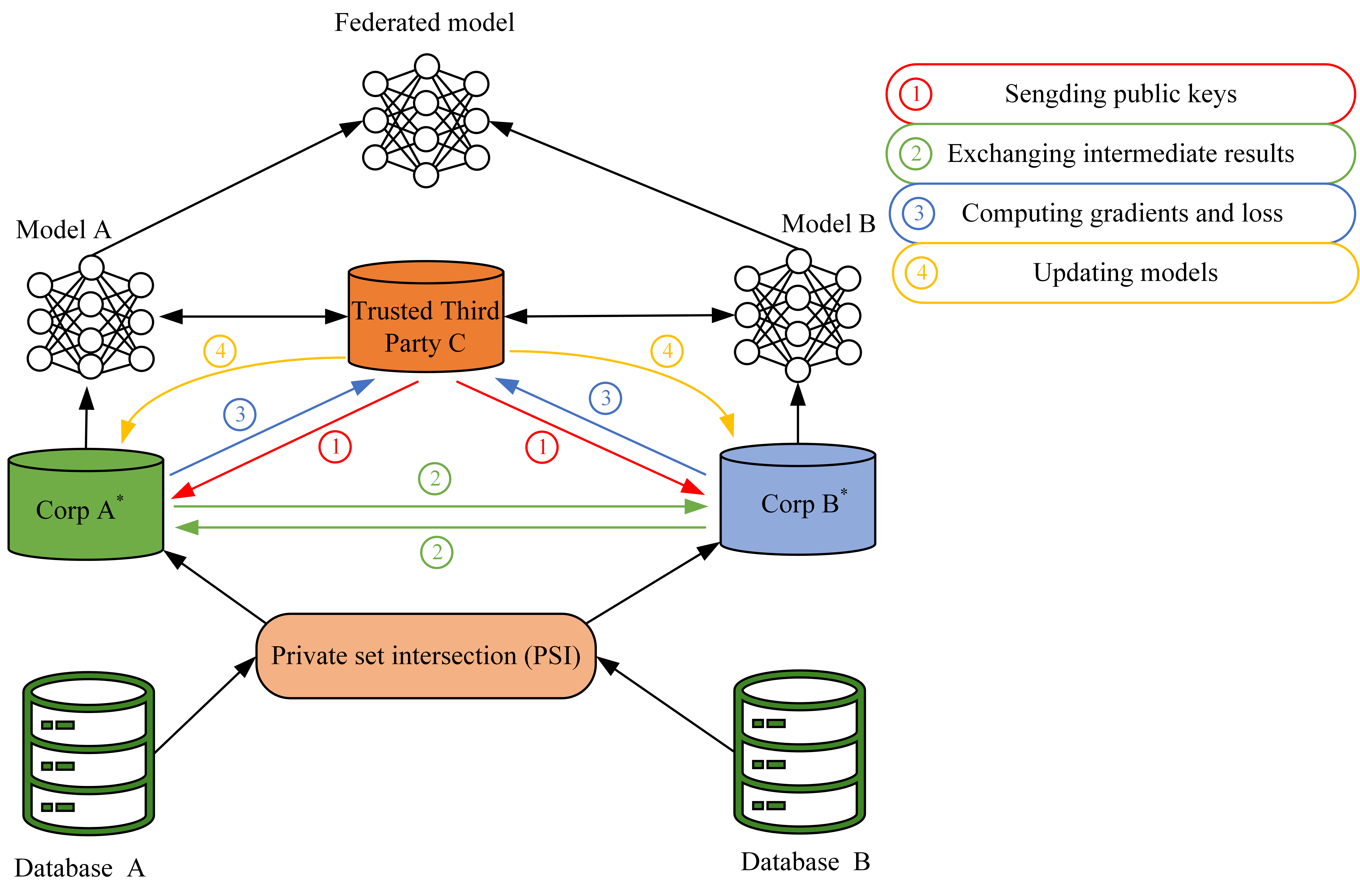}
\caption{Vertical  federated learning framework.}
\label{fig_5}
\end{figure}

 VFL is applicable to the case where the data samples of the federated learning participants overlap more, i.e., the data samples are aligned between the participants, but they differ in data features. It is formally similar to the case of dividing the data in a table vertically. Therefore, VFL is known as feature-divided federated learning\cite{DBLP:journals/ftml/KairouzMABBBBCC21}, its framework is shown in Figure \ref{fig_5}.
\par VFL is often used in Cross-Silo scenarios, and current linear models (such as linear regression, logistic regression, etc.), boosted tree models, SecureBoost\cite{DBLP:journals/expert/ChengFJLCPY21}, neural networks, vertical matrix decomposition in personalized recommendation, and vertical factorization machines have all been implemented on VFL. The paper\cite{DBLP:journals/corr/abs-1711-10677} propose a VFL scheme to train a privacy-preserving logistic regression model. Xu et al. propose a federated deep Learning framework\cite{2020Privacy} to reduce the negative impact of irregular users on the training accuracy.

\subsection{Full Homomorphic Encryption} 
The concept of Homomorphic Encryption (HE) was first proposed by Rivest et al. in 1978\cite{RonaldRivest}. The term "homomorphic" means that after the data is homomorphically encrypted, the result of the computation on the ciphertext is the same after decryption as the result of the direct computation on the plaintext. 
\par Since the concept of HE was proposed, researchers have proposed many encryption algorithms that support certain functions of ciphertext computation, including RSA algorithm, Paillier and so on. These algorithms are regarded as Partial Homomorphic Encryption (PHE).
\par It should be noted that Paillier algorithm is a core module of Federated Learning, which is proposed by Pascal Paillier in 1999\cite{DBLP:conf/eurocrypt/Paillier99}. It is a typical PHE algorithm that supports ciphertext addition, ciphertext-plaintext multiplication, and does not support ciphertext-ciphertext multiplication and complex exponential and logarithmic operations.
\par In 2009, Gentry  gave the first theoretically feasible blueprint for FHE algorithms \cite{STOC:gen09},  since then great progress have been achieved in designing FHE algorithms. Up to now, there are two types of FHE structures of FHE algorithms that are considered to be practical. One is based on the work of Brakerski et al. in 2011 \cite{DBLP:conf/crypto/Brakerski12,brakerski2014efficient,ITCS:BraGenVai12}, which is good at dealing with numerical operations, the other is based on the structure proposed by  Gentry, Sahai and Waters in 2013 \cite{DBLP:conf/crypto/GentrySW13} which performs well in logical operations. In this paper, we choose  FHE algorithm named CKKS which is  proposed by Cheon et al. in 2017\cite{DBLP:conf/asiacrypt/CheonKKS17}, it is known as the most efficient FHE algorithm for numerical operations because it uses a novel embedding technique to improve computation efficiency at the cost of the accuracy of outputs. However, we notice that this feature makes the algorithm  extremely fit for federated learning because we can convert these output errors  into a protection strategy against gradient attacks and make   federated learning algorithms more secure and efficient at the same time.

\subsection{Private Set Intersection}
\par Privacy Set Intersection (PSI) comprises a category of application-specific techniques with extensive potential in the realm of privacy computing. Currently, PSI is undergoing rapid development and has emerged as one of the more pragmatic security computing methods, finding applications in private address book finding\cite{DBLP:journals/iacr/DemmlerRRT18}, online advertisement real effect computation\cite{DBLP:journals/fgcs/LvYYCFLLLLZ20}, genetic sequence match detection\cite{DBLP:conf/bibm/Shen0WFD18}, and beyond. 
\par Numerous classifications of privacy set intersection (PSI) protocols exist, with the primary categorization based on the underlying cryptography. This includes public key cryptography-based PSI schemes, oblivious transfer-based PSI schemes, generic MPC-based PSI schemes, and homomorphic encryption-based PSI schemes. 
 \par The first PSI protocol, developed by Meadows \cite{DBLP:conf/sp/Meadows86}, was founded on public key encryption, harnessing the multiplicative homomorphic properties of the Diffie-Hellman key exchange. Aranha\cite{DBLP:conf/ccs/AranhaLO022} introduced a novel two-party PSI protocol designed to minimize the sender's overhead from both theoretical and practical perspectives. Dong et al.\cite{DBLP:conf/ccs/DongCW13} innovatively incorporated Bloom filters into PSI for the first time and, when combined with OT extensions, expanded the PSI protocol's capacity to handle sets to levels surpassing a billion for the first time. Chase et al.\cite{DBLP:conf/crypto/ChaseM20} proposed a new lightweight multi-point oblivious pesudorandom function (OPRF) protocol based on oblivious transfer (OT) extension which achieves a better balance between computation and communication than existing PSI protocols.

\section{Federated Learning with SecureBoost based on FHE}
\setlength{\parskip}{0.5em}
\subsection{Classical SecureBoost Model based on PHE}
\subsubsection{XGBoost Model}
\par Boosting tree is an efficient and widely used machine learning method that performs well in many machine learning tasks due to its high efficiency and interpretability. For example, XGBoost\cite{2016XGBoost} has been widely used in a variety of applications, including credit risk analysis and user behavior studies. 
\par Given a dataset $\mathbf{X} \in \mathbb{R}^{n\times d}$, denoting $n$ samples and $d$ features, XGBoost predicts the results using $K$ regression trees:
\[\hat{y}_{i}=\sum_{k=1}^{K} f_{k}(\mathbf{x}_{i}), \ \mathbf{x}_{i} \in \mathbf{X}.\]
\par In order to learn $f_{k}$ as described above, at round $t$, XGBoost uses a greedy approach to learn the model $f_{t}$ by minimizing the following objective:
\[\mathcal{L} ^{(t)}\simeq \sum_{i=1}^{n} [l(y_{i} ,\hat{y} ^{(t-1)}) + g_{i}f_{t}(x_{i}) + \frac{1}{2}h_{i}f_{t}^{2}(x_{i})] + \Omega (f_{t})   ,\]
where $\Omega (f_{t})=\gamma T+\frac{1}{2}\lambda ||w||^{2}$, $g_{i}=\partial_{\hat{y}^{(t-1)}}l(y_{i},\hat{y}^{(t-1)}) $ and $h_{i}=\partial _{\hat{y}^{(t-1)}}^{2}l(y_{i},\hat{y}^{(t-1)})$.

During the $t$-th round of constructing the decision tree, the model splits from nodes with depth 0 until it reaches the maximum depth. The optimal splitting node is given according to the following equation (where $I_{L}$, $I_{R}$ denote the data of the left and right nodes after splitting):
\[\mathcal{L}_{split}=\frac{1}{2}[\frac{(\sum _{i\in I_{L}}g_{i})^{2} }{\sum _{i\in I_{L}}h_{i}+\lambda}+\frac{(\sum _{i\in I_{R}}g_{i})^{2} }{\sum _{i\in I_{R}}h_{i}+\lambda}-\frac{(\sum _{i\in I}g_{i})^{2} }{\sum _{i\in I}h_{i}+\lambda}]-\gamma.\]
\par After constructing the optimal tree structure (splitting result), the weight $w_{j}^{*} $ of leaf node $j$ can be calculated according to the following equation, where $I_{j}$ is the data on leaf $j$:
\[w_{j}^{*}=-\frac{\sum _{i\in I_{j}}g_{i}}{\sum _{i\in I_{j}}h_{i}+\lambda } .\]
\subsubsection{SecureBoost Model}
Cheng et al. proposed an end-to-end privacy-preserving boosting tree algorithmic framework, called SecureBoost, to implement machine learning in a federated environment\cite{DBLP:journals/expert/ChengFJLCPY21}.
\par Two types of parties involved are defined in SecureBoost, namely active party and passive party.
\par \textbf{Active Party:} The active party is defined as a data provider that possesses both the data matrix and the class labels. Since class labeling information is essential for supervised learning, it is natural for the active party to assume the responsibility of being the dominant server in federated learning.
\par \textbf{Passive Party:} A data provider with only a data matrix is defined as a passive party. The passive party plays the role of a client in a federated learning environment.
\par From XGBoost, it can be seen that: (1) The computation of segmentation candidate nodes and leaf optimal weights depends only on $g$ and $h$, so it is easy to adapt to the federated learning setting. (2) Label can be inferred from $g_{i}$ and $h_{i}$, so it is protected.
\par Based on the above notions, the federated gradient boosting tree algorithm is obtained. According to (1) above, it can be seen that the passive party can use only their local data and $g_{i}$, $h_{i}$ to determine their locally optimal segmentation. However, according to (2) above, $g_{i}$ and $h_{i}$ should be regarded as sensitive data because they are able to disclose class labeling information to the passive party. Therefore, to keep $g_{i}$ and $h_{i}$ confidential, the active party needs to encrypt $g_{i}$ and $h_{i}$ before sending them to the passive party. The remaining problem is how to determine the local optimal split for each passive party.
\par According to the formula $\mathcal{L}_{split}$ for computing the optimal split, the optimal split can be found if $g_{l}=\sum _{i\in I_{L}}g_{i}$, $h_{l}=\sum _{i\in I_{L}}h_{i}$ are computed for each possible split. 
\par The additive homomorphic encryption scheme Paillier is used in SecureBoost. Denote the Paillier ciphertext of the data $u$ as $<u>$, the Paillier system has the following property because it is homomorphic for addition: $<u>+<v>=<u+v>$. Hence there is: $<h_{l}>=\sum _{i\in I_{L}}<h_{i}>$, $<g_{l}>=\sum _{i\in I_{L}}<g_{i}>$. Therefore, the optimal splits can be found as follows: the passive party first computes all the possible splits locally, computes $<g_{l}>$, $<h_{l}>$, and then sends them to the active party. Then the active party decrypts and calculates the optimal split node according to the optimal split formula. SecureBoost uses the approximation algorithm utilized in XGBoost so that it does not have to exhaust all possible values and can be optimized in some way. The flow of details is shown in Algorithm 1.
\begin{table}[]
\begin{tabular}{l}
\hline
{\textbf{Algorithm 1} Aggregate Encrypted Gradient Statistics using Paillier} \\ \hline
\textbf{Input}: I, instance space of current node                                  \\
\textbf{Input:} d, feature dimension                                               \\
\textbf{Input:} $\{<g_{i} >, <h_{i} >\}_{i\in I}$          \\
\textbf{Output:} $\mathbf{G} \in \mathbb{R}^{d\times l}, \mathbf{H} \in \mathbb{R}^{d\times l}$                                                              \\
\quad 1: \textbf{for} $k=0 \to d$ \textbf{do}                                             \\
\quad 2: \quad \quad Propose $S_{k}=\{s_{k1},s_{k2},...,s_{kl}\} $ by percentiles on feature $k$                          \\
\quad 3: \textbf{end for}                                                                \\
\quad 4: \textbf{for} $k=0 \to d$ \textbf{do}                                             \\
\quad 5: \quad \quad $\mathbf{G}_{kv}=\sum _{i\in \{i|s_{k,v}\ge x_{i,k}> s_{k,v-1}\}<g_{i} >} $                                                                   \\
\quad 6: \quad \quad $\mathbf{H}_{kv}=\sum _{i\in \{i|s_{k,v}\ge x_{i,k}> s_{k,v-1}\}<h_{i} >} $                                                                   \\
\quad 7: \textbf{end for}                                                                \\ \hline
\end{tabular}
\end{table}
\subsection{Federated Learning with SecureBoost based on FHE}
In Algorithm 1 above, the partially homomorphic encryption algorithm Paillier is used to compute $<g_{l}>$, $<h_{l}>$ and prevent label from leakage at the same time, these goals can be achieved by FHE with better efficiency. Therefore, we use fully homomorphic encryption algorithm CKKS to replace Paillier algorithm to obtain Algorithm 2. In CKKS, denote the ciphertext of the data $u$ as $[[u]]$, the following properties are available: $[[u]]+[[v]]=[[u+v]], [[u]]\times [[v]]=[[u\times v]]$.
\begin{table}[]
\begin{tabular}{l}
\hline
{\textbf{Algorithm 2} Aggregate Encrypted Gradient Statistics using CKKS} \\ \hline
\textbf{Input}: I, instance space of current node                                  \\
\textbf{Input:} d, feature dimension                                               \\
\textbf{Input:} $\{[[g_{i} ]], [[h_{i} ]]\}_{i\in I}$          \\
\textbf{Output:} $\mathbf{G} \in \mathbb{R}^{d\times l}, \mathbf{H} \in \mathbb{R}^{d\times l}$                                                              \\
\quad 1: \textbf{for} $k=0 \to d$ \textbf{do}                                             \\
\quad 2: \quad \quad Propose $S_{k}=\{s_{k1},s_{k2},...,s_{kl}\} $ by percentiles on feature $k$                          \\
\quad 3: \textbf{end for}                                                                \\
\quad 4: \textbf{for} $k=0 \to d$ \textbf{do}                                             \\
\quad 5: \quad \quad $\mathbf{G}_{kv}=\sum _{i\in \{i|s_{k,v}\ge x_{i,k}> s_{k,v-1}\}[[g_{i} ]]} $                                                                   \\
\quad 6: \quad \quad $\mathbf{H}_{kv}=\sum _{i\in \{i|s_{k,v}\ge x_{i,k}> s_{k,v-1}\}[[h_{i} ]]} $                                                                   \\
\quad 7: \textbf{end for}                                                                \\ \hline
\end{tabular}
\end{table}
\par 
\subsubsection{Approximate Algorithm for Split Finding}
In Algorithm 2, it is necessary to find the set of candidate split points $\{s_{k1},s_{k2},...,s_{kl}\}$, and a ranking function is used in XGBoost to find the set of candidate split points.
Given a set $D_{k}=\{(x_{1k},h_{1}),(x_{2k},h_{2}),...,(x_{nk},h_{n})\} $ consisting of the $k$-th feature of a sample in a dataset and the second-order derivatives of the sample points on the loss function. Subsequently use the percentage of the data distribution to define a ranking function $r_{k}:R\longrightarrow [0,1) $.
\[r_{k}(z)=\frac{1}{\sum _{(x,h)\in D_{k}}h}\sum _{(x,h)\in D_{k},x<z}h .\]
\par This ranking function represents the proportion of samples with values of feature $k$ less than $z$ out of the total samples that the candidate split points need to satisfy:
\[|r_{k}(s_{k,j})-r_{k}(s_{k,j+1})|< \epsilon ,s_{k1}=min_{i}\mathbf{x}_{ik},s_{kl}=max_{i}\mathbf{x}_{ik}, \]
then we will get $\frac{1}{\epsilon}$ candidate splitting points.
\par The above method uses the second order derivatives $h$ as weights to divide the samples in which comparison operations are used. However, after using FHE, $h$ as a ciphertext is passed from the active party to the passive party, in which the homomorphic ciphertext comparison operation is complicated. 
\par Therefore we find the split point based on the distribution of features and divide the sample under the current node into $\frac{1}{\epsilon}$ parts, thus getting $\frac{1}{\epsilon}$ candidate split points. Experiments show that this method of finding candidate split points does not reduce the accuracy of the model, but rather makes the model have better efficiency.
\par
\begin{figure}[!h]
\centering
\includegraphics[width=3.4in]{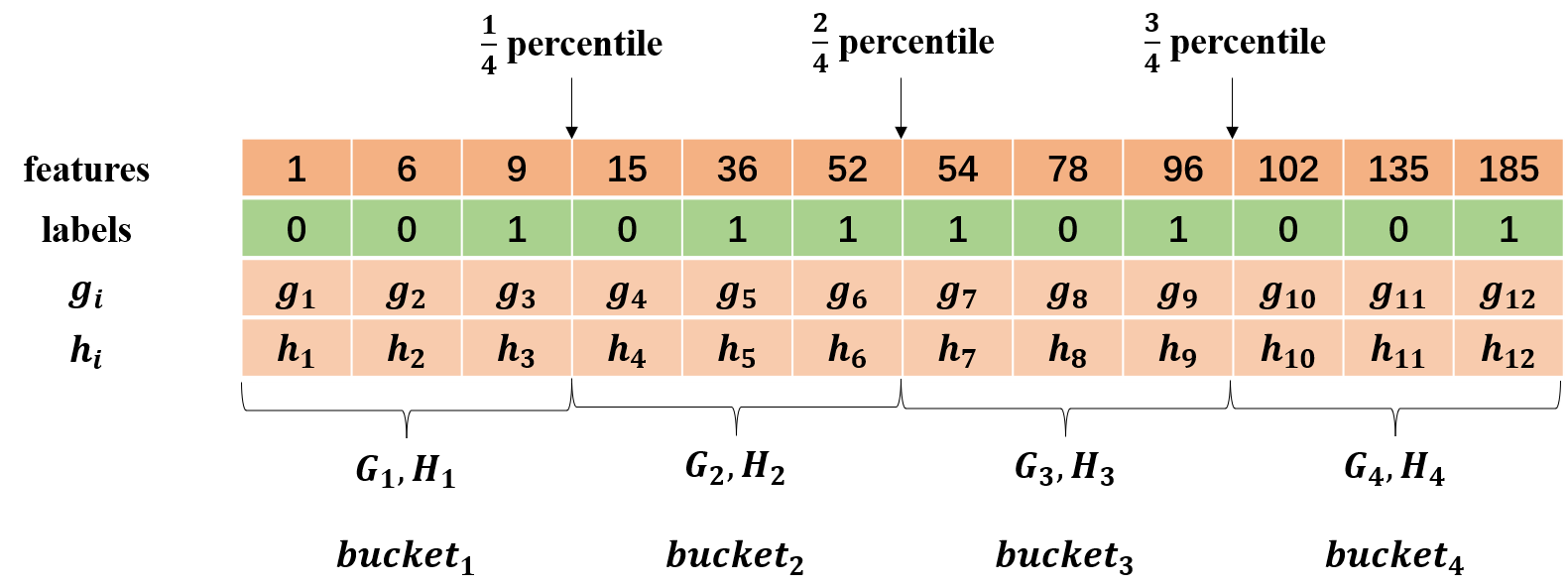}
\caption{Schematic diagram for finding candidate split points.}
\label{percentile}
\end{figure}

 Figure \ref{percentile} is a schematic diagram to visualize how Algorithm 2 finds the candidate split points. Firstly, according to the distribution of the features, find the quartile point, which divides the samples under the current node into 4 parts; then according to the loss function, get the first-order derivatives $g_{i}$ and second-order derivatives $h_{i}$ of each sample in a certain iteration process; then sum and summarize the first-order derivatives and second-order derivatives of the samples in each interval, respectively, to get $(G_{1},H_{1}),(G_{2},H_{2}),(G_{3},H_{3}),(G_{4},H_{4})$; finally, using the greedy algorithm, search for the optimal splitting point of the intervals, because there are only 4 intervals, so there are only 3 combinations.
\subsubsection{Ciphertext packing}
In Algorithm 2, it is necessary to use the encrypted $g_{i}$ and $h_{i}$ to compute the aggregated gradient information for each feature. By observation, it can be seen that $g_{i}$ and $h_{i}$ enjoy the same operation, i.e., homomorphic addition operation. Whereas CKKS provides ciphertext packing techniques to encrypt multiple plaintexts into a single ciphertext, then doing homomorphic operations on such ciphertexts is equivalent to performing the same operations on the plaintexts on each component in parallel.
Thus, using this technique, we can pack $g_{i}$ and $h_{i}$ into a ciphertext vector. This operation will reduce the cost of encryption, decryption, and homomorphic addition operations, and greatly improve the efficiency of model training.
\subsubsection{Subtraction between tree nodes}
Above, it was mentioned how to get the candidate split points of the features and then aggregate $g_{i}$ and $h_{i}$. This process is equivalent to building buckets for the features and later dividing the samples into different buckets.
\par
\begin{figure}[!h]
\centering
\includegraphics[width=3.4in]{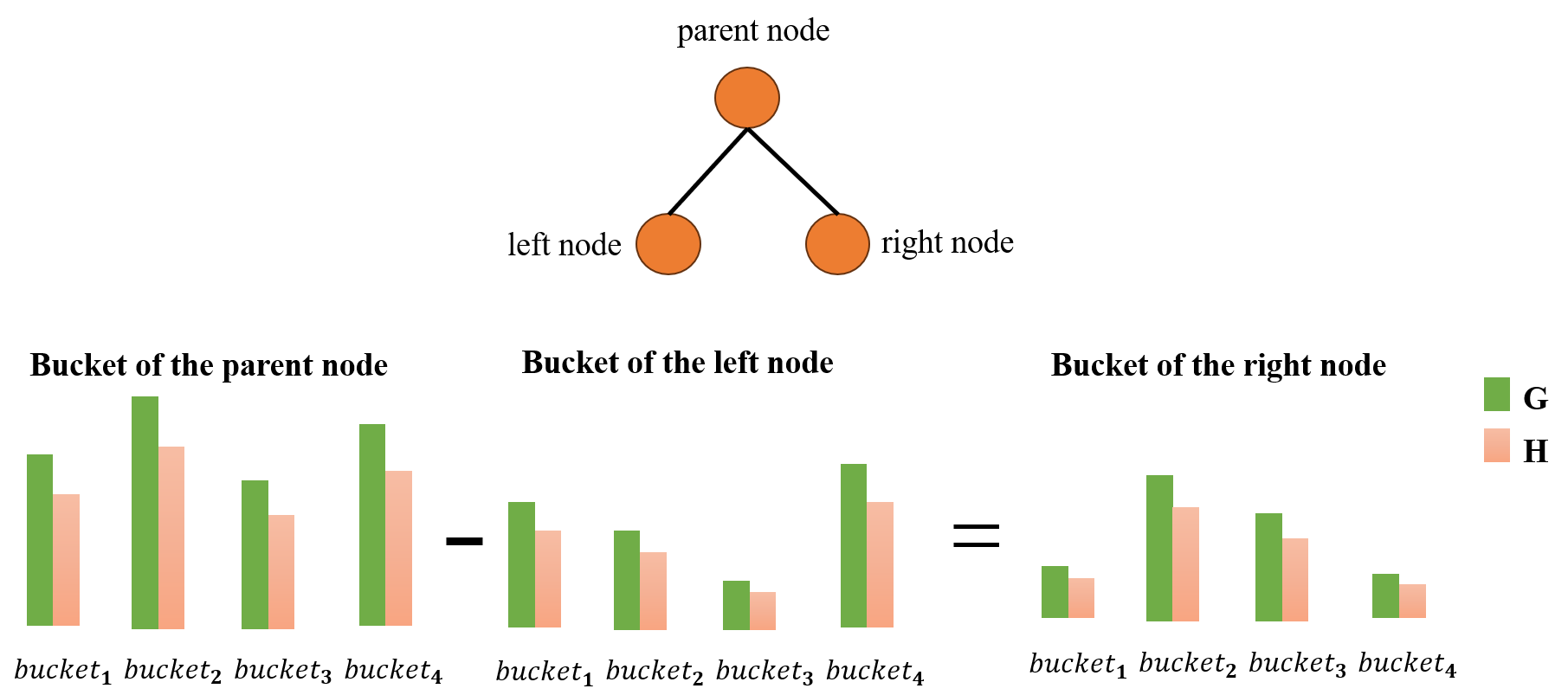}
\caption{Subtraction between tree nodes.}
\label{sub}
\end{figure}

 In the context of a specific parent node within the tree, when the node is split, the samples associated with this node are distributed either to the left node or the right node. Subsequently, for each feature, the sum of the buckets for the left child node and the right child node must equate to the bucket of the parent node after the split. Leveraging this property, as shown in Figure \ref{sub}, we can focus on constructing buckets for the parent node and the left child node, as the bucket for the right child node can be derived by subtracting the bucket of the left child node from the bucket of the parent node.
\subsection{Federated Inference based on the Learned Model}
\subsubsection{Classic federated inference}
SecureBoost \cite{DBLP:journals/expert/ChengFJLCPY21} describes how to classify new instances using a learned model (distributed between parties), even if the features of the instances to be classified are private and distributed between parties. Since each party knows only its own features but nothing about the other parties, a secure distributed inference protocol is needed.
\par
\begin{figure}[!h]
\centering
\includegraphics[width=3.4in]{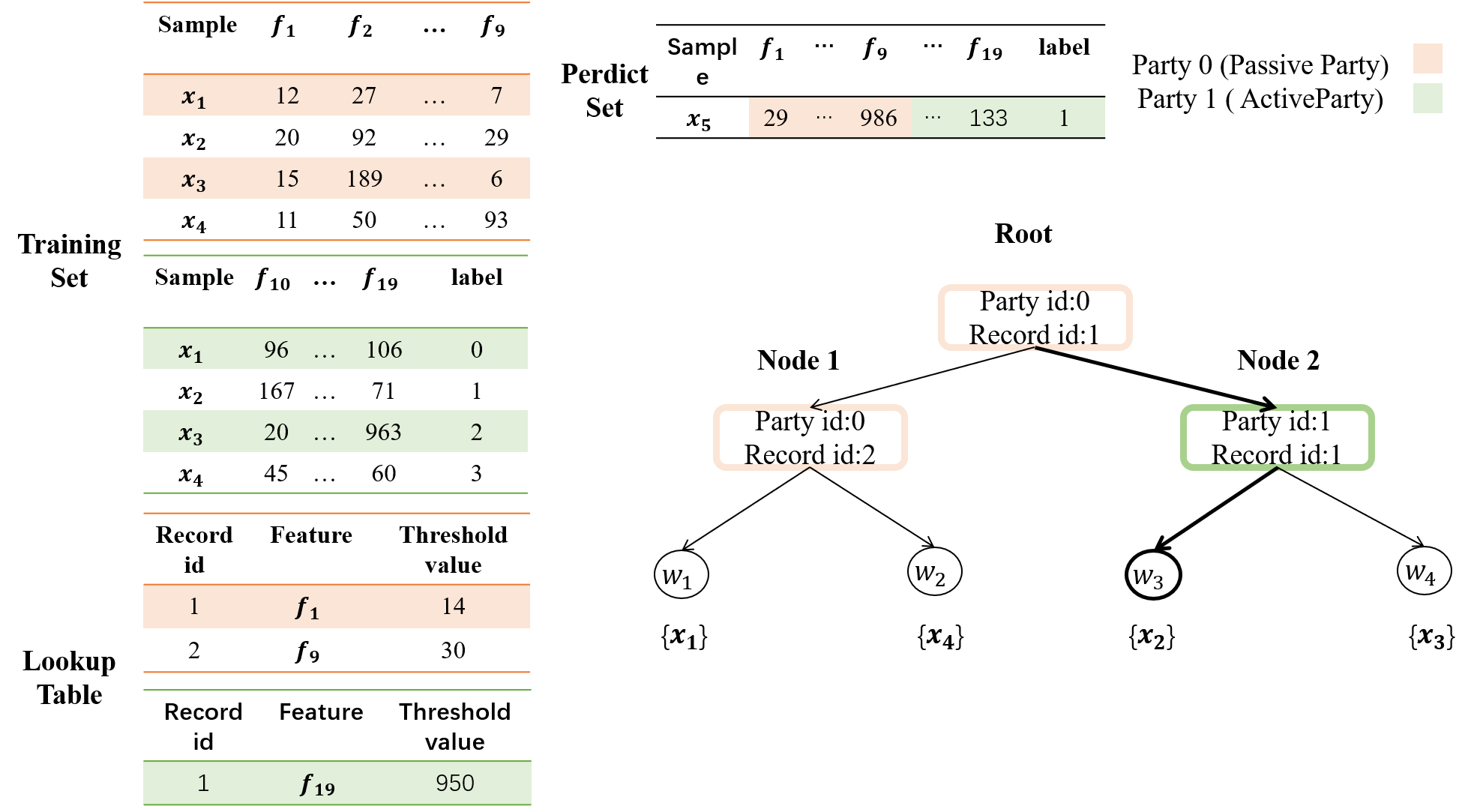}
\caption{An illustration of classic federated inference.}
\label{predict}
\end{figure}

 Now consider a two-party system as shown in Figure \ref{predict}. Specifically, the first party is a passive party with features $f_{0}, f_{1}, ..., f_{9}$. The second party is the active party, which owns the features $f_{10}, f_{11}, ..., f_{19}$ as well as the label $y$ and the learned tree model. Suppose that for a new sample $x_{5}$, it is desired to predict the label $y$ of the sample, then all the involved parties must collaborate in making the prediction. The whole process is coordinated by the active party. Starting from the root node of the tree, the active party knows that the passive party 0 holds the root node through the node record $[party\ id:0,\ record \ id:1]$, thus requiring party 0 to retrieve the corresponding feature from its lookup table based on $record id:1$. Since the classification feature of this node is $f_{1}$ and the party 0 knows that $f_{1}$ of sample $x_{5}$ is $29$, which is greater than the threshold $14$, it decides that it should move down to its right child node, Node 2. The active party then retrieves its own lookup table by referring to the record $[party\ id:1,\ record\ id:1]$ associated with Node 2. This process continues until the leaf is reached.
\subsubsection{Improved federated inference using PSI}
\begin{figure}[!h]
\centering
\includegraphics[width=3.4in]{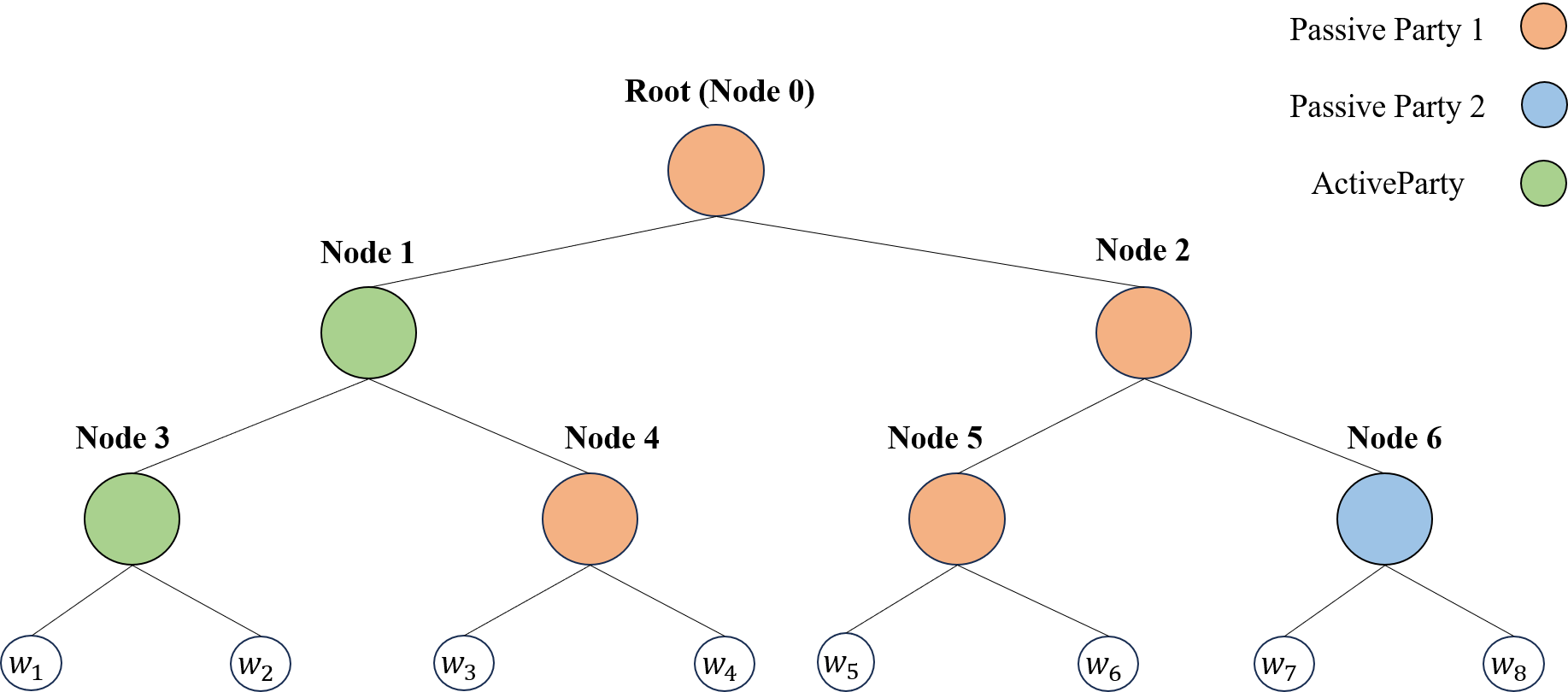}
\caption{Schematic diagram of the division path of the sample for three participants.}
\label{path1}
\end{figure}

\begin{figure}[!h]
\centering
\includegraphics[width=3.4in]{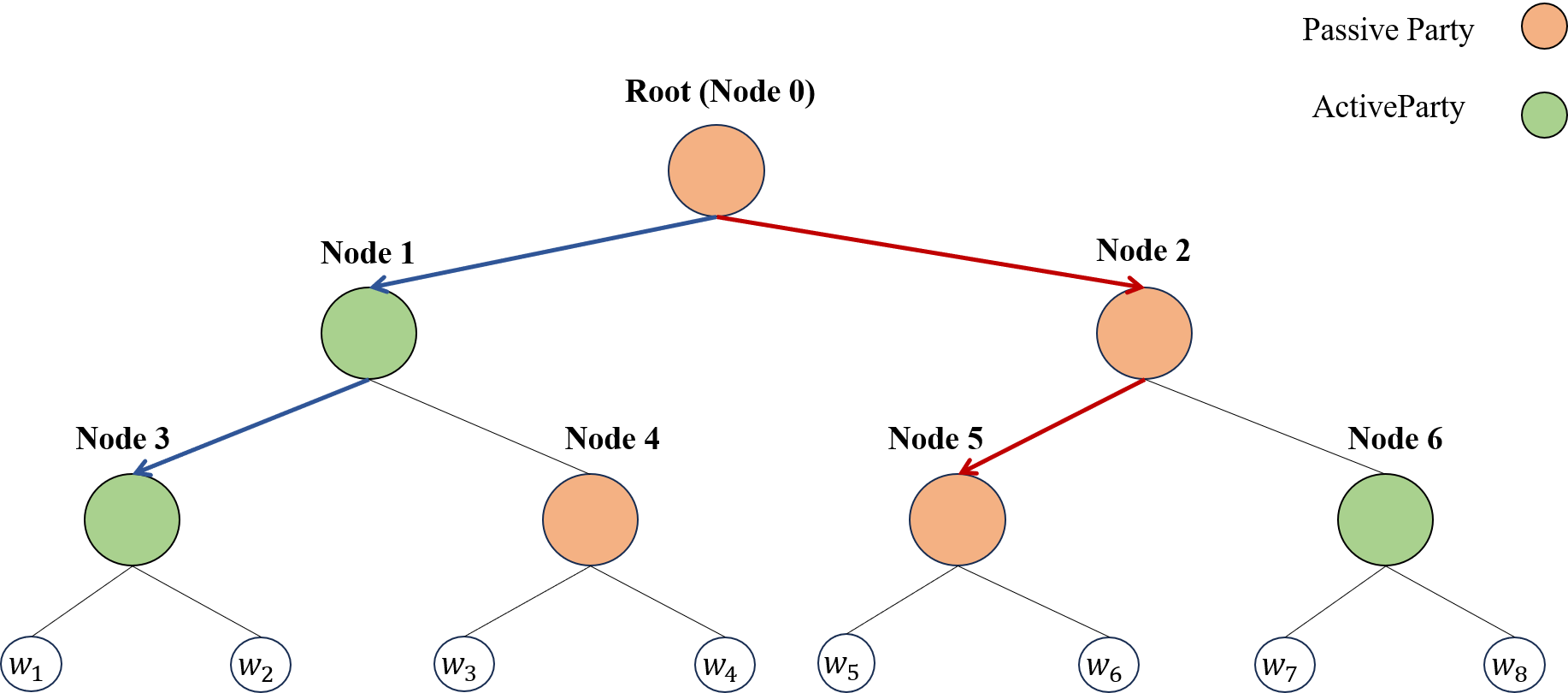}
\caption{Schematic diagram of the division path of the sample for two participants.}
\label{path}
\end{figure}

As can be seen from Figure \ref{predict}, the number of interactions between active and passive parties increases with the depth of the tree in the federated inference process, leading to a corresponding increase in communication. This means that the number of interactions between active and passive parties, as well as the amount of communication, varies throughout the federated inference process. This variation occurs because the path from the root node to the leaf nodes differs for each sample, resulting in a varying number of nodes assigned to the passive party along this path. Consequently, in certain scenarios, the passive party can use the number of interactions with the active party and the amount of communication to determine the sample's division path and its associated leaf node, and which leaf node the sample belongs to is sensitive information, and there will be security problems if it is leaked. Consider a tree model depicted in Figure \ref{path1} involving three or more participants, a passive party can discern the final result by simply eavesdropping on communications between the active party and other parties, even if they aren't directly engaged in the division process. For instance, if passive party 2 monitors the communication between the active party and passive party 1 without being involved in the sample division process, the sample's division path is Root $\longrightarrow$ Node 1 $\longrightarrow$ Node 3 after one communication between active and passive party 1, Root $\longrightarrow$ Node 1 $\longrightarrow$ Node 4 after two communications between active and passive party 1, and Root $\longrightarrow$ Node 2 $\longrightarrow$ Node 5 after three communications between them. Additionally, in the two-participant scenario illustrated in Figure \ref{path}, if the number of communications between the passive party and the active party is 1, then the division path of the sample is: Root $\longrightarrow$ Node 1 $\longrightarrow$ Node 3; if the number of communications between the passive party and the active party is 3, then the division path of the sample is: Root $\longrightarrow$ Node 2 $\longrightarrow$ Node 5.
\par In both scenarios, apart from the active party, the passive party possesses the capability to infer the final division outcome by analyzing the sample's division pathway. In order to solve this problem, a method that requires only one communication between the active and passive parties to complete the federated inference is needed. That is, the method to be introduced next uses PSI in the process of federated inference. This method ensures that the final division of the sample cannot be surmised by the passive party except for the active party.


\par
\begin{figure}[!h]
\centering
\includegraphics[width=3.4in]{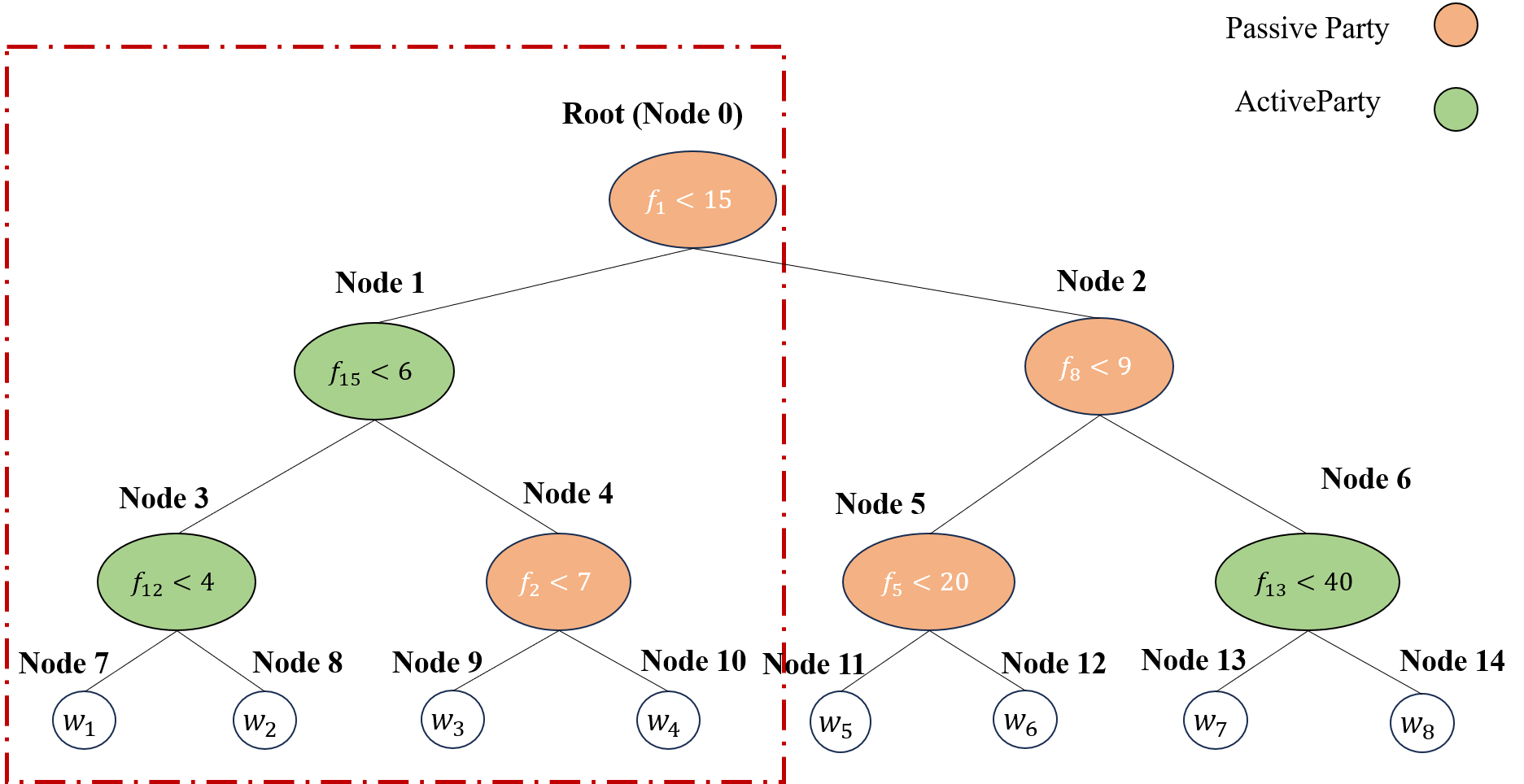}
\caption{An illustration of federated inference using PSI.}
\label{psi}
\end{figure}

 The tree model is shown in Figure \ref{psi}, where the tree nodes are numbered for ease of illustration and the thresholds of the active and passive party nodes are shown directly (but the thresholds are only visible to themselves). Again assuming that the passive party has features $f_{0}, f_{1}, ..., f_{9}$, and the active party owns features $f_{10}, f_{11}, ..., f_{19}$, and the selected features in the tree model are $f_{1}, f_{15}, f_{8},f_{12}, f_{2}, f_{5}, f_{13}$. The existing sample $x_{p}$ is distributed on two sides, the passive side: $f_{1}=13, f_{2}=10, f_{5}=30, f_{8}=10$, active side: $f_{12}=5, f_{13}=50, f_{15}=10$.
\par In simple terms, the participating parties will remove the nodes they can judge from the list of nodes in order, and then the remaining nodes will be subjected to the PSI operation. For example, for the passive party to start having the node list [0, 1, 2, 3, 4, 5, 6, 7, 8, 9, 10, 11, 12, 13, 14], the sample $x_{p}$'s $f_{1}=13$ is less than the threshold 15 of Node 0 so the sample will be classified as Node 1, so the passive party will remove 2, 5, 6, 11, 12, 13, and 14 from the node list; the sample $x_{p}$'s $f_{2}=10$ is greater than the Node 4's threshold 7, so the passive side will remove 9 from the node list. eventually, the passive side node list remains [0, 1, 3, 4, 7, 8, 10].
\par For the active side start also has node list [0, 1, 2, 3, 4, 5, 6, 7, 8, 9, 10, 11, 12, 13, 14], sample $x_{p}$'s $f_{15}=10$ is greater than the threshold 6 for Node 1, so the active side removes 3, 7, and 8 from the node list; sample $x_{p}$'s $f_{13}=50$ is greater than the threshold 40 for Node 6, so the active side removes 13. Eventually, the passive side node list remains [0, 1, 2, 4, 5, 6, 9, 10, 11, 12, 14].
\par Then active and passive side perform PSI operation on the final list of nodes and finally active side gets the PSI result [0, 1, 4, 10] and hence knows that the sample $x_{p}$ is finally classified to the leaf node $w_{4}$. In a scenario involving n participants, each passive party conducts an individual PSI operation with the active party, necessitating n-1 PSI operations to accomplish the aforementioned inference process.

\par The advantage of using PSI is that regardless of the feature values of a sample, a sample only needs to perform PSI once to complete the federated inference process, avoiding the problem of deducing the division of a sample by the number of communications.
\par According to the recent results, the PSI protocol based on multipoint inadvertent pseudo-random function construction proposed by Chase et al.\cite{DBLP:conf/crypto/ChaseM20} is the current practical solution with excellent computational performance and communication overhead, so we implement our algorithm's PSI protocol based on it.

\subsection{Performance Analysis} 
\subsubsection{Model training phase}
\hfill\par (1) Application \uppercase\expandafter{\romannumeral1}: Cancer prediction(medical field) 
\par \textbf{Dataset: }This experiment uses the breast cancer dataset \cite{Federated-Learning}, which contains 569 samples with 30 features. The data samples are derived from measurements of the breast lump image and whether it is cancerous or not, and the goal is to use these measurements to predict whether the lump is cancerous or not. In this dataset, the number of positive and negative samples is balanced, so the model quality can be assessed using accuracy. 
\par \textbf{Experimental environment: }Intel(R) Xeon(R) Gold 6226R CPU @ 2.90GHz
\par
\begin{figure}[!h]
\centering
\includegraphics[width=3.4in]{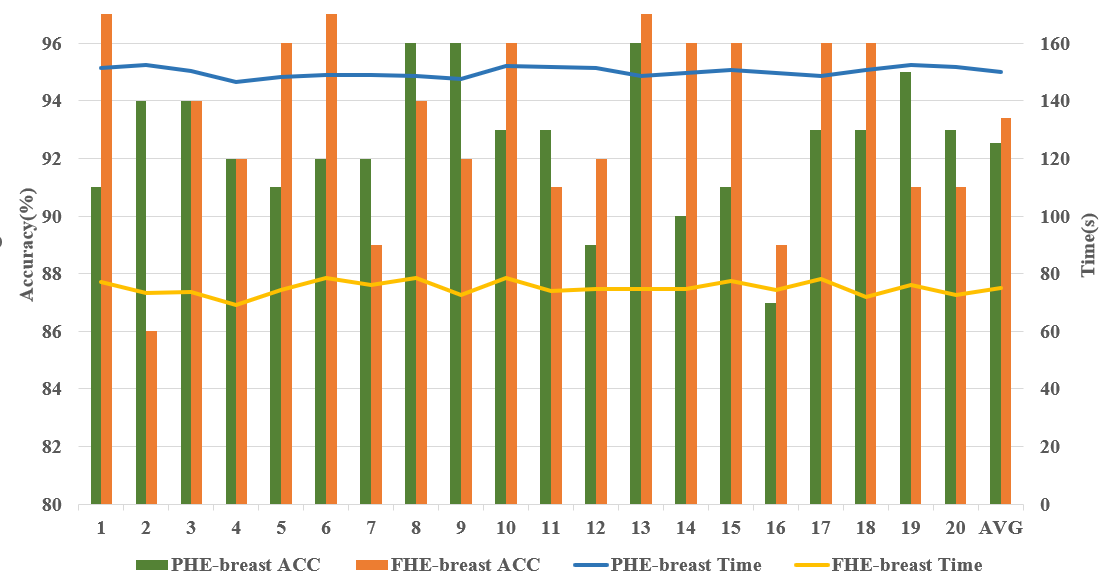}
\caption{Comparison of efficiency and accuracy in breast cancer dataset under secureboost.}
\label{secure boost breast}
\end{figure}

 \textbf{Experimental phase: }Under the SecureBoost model based on FHE, we were conducted 20 experiments, the model training time (excluding communication time consumption) and the accuracy of the model are obtained respectively. The control group used the classical SecureBoost Model based on PHE propsed in \cite{DBLP:journals/expert/ChengFJLCPY21}, and the results obtained under the same dataset are shown in Figure \ref{secure boost breast}.

\par \textbf{Result analysis: }By analyzing the experimental data, we can see that the average training time of the SecureBoost model based on FHE is 75.04s, and the average accuracy is 93.40\%; while the average training time of the classical SecureBoost model with the same dataset is 150.07s, and the average accuracy is 92.55\%. Thus, the training efficiency of our scheme is 2 times higher than that of classical SecureBoost model in this dataset and the accuracy also improved slightly, with an average range of 0.85\%.

\hfill\par (2) Application \uppercase\expandafter{\romannumeral2}: Wholesale customers prediction (business field) 
\par \textbf{Dataset: }The data set \cite{misc_wholesale_customers_292} refers to clients of a wholesale distributor. It includes the annual spending in monetary units (m.u.) on diverse product categories.
\par \textbf{Experimental environment: }Intel(R) Xeon(R) Gold 6226R CPU @ 2.90GHz
\par
\begin{figure}[!h]
\centering
\includegraphics[width=3.4in]{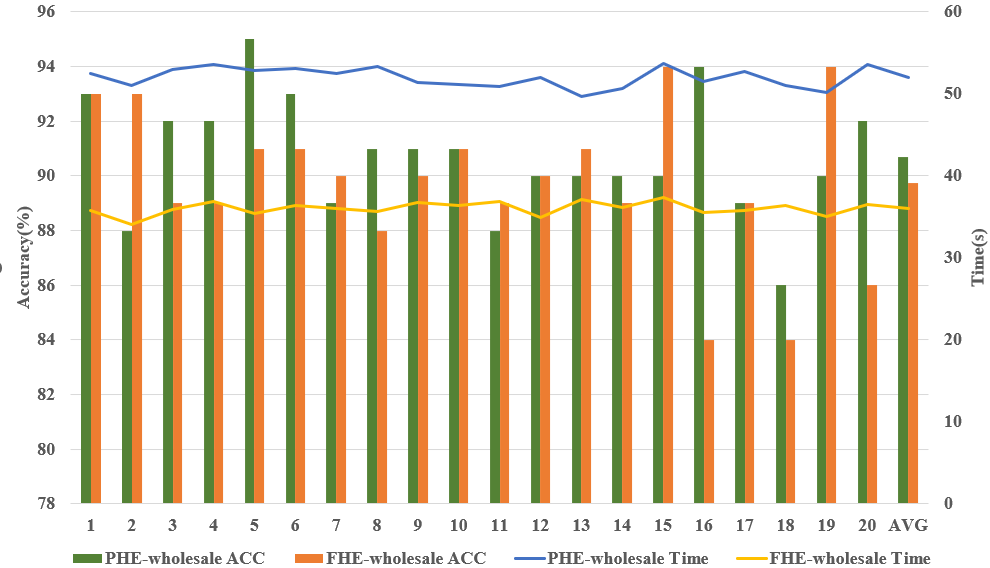}
\caption{Comparison of efficiency and accuracy in Wholesale customers dataset under secureboost.}
\label{secure boost Wholesale}
\end{figure}

 \textbf{Experimental phase: }Under the SecureBoost model based on FHE, we were conducted 20 experiments, the model training time (excluding communication time consumption) and the accuracy of the model are obtained respectively. The control group used the classical SecureBoost Model based on PHE propsed in \cite{DBLP:journals/expert/ChengFJLCPY21}, and the results obtained under the same dataset are shown in Figure \ref{secure boost Wholesale}.
\par \textbf{Result analysis: }By analyzing the experimental data, we can see that the average training time of the SecureBoost model based on FHE is 36.03s, and the average accuracy is 89.75\%; while the average training time of the classical SecureBoost model with the same dataset is 52.01s, and the average accuracy is 90.70\%. Thus, the training efficiency of our scheme is 1.4 times higher than that of classical SecureBoost model in this dataset and the accuracy decreased slightly, with an average range of 0.95\%.

\subsubsection{Federated inference phase}
\hfill\par After training the tree model, we introduce a novel federated inference approach using Private Set Intersection (PSI). This method not only enhances security but has also been empirically demonstrated to reduce communication overhead in specific scenarios. The experiment is conducted with the breast cancer dataset.
\par \textbf{Experimental environment: }Intel(R) Xeon(R) Gold 6226R CPU @ 2.90GHz
\par
\begin{figure}[!h]
\centering
\includegraphics[width=3.4in]{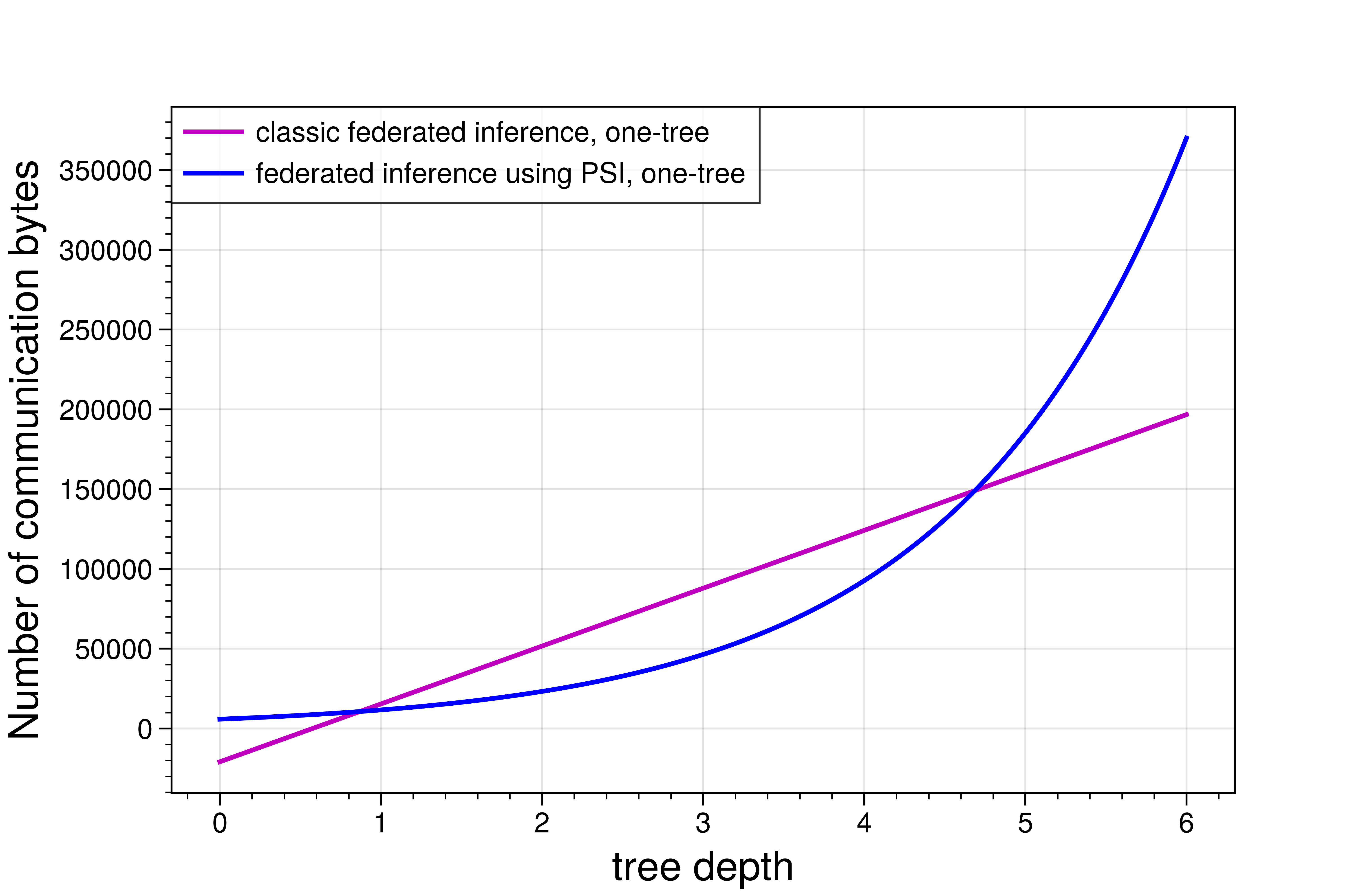}
\caption{Comparison of communications under one tree.}
\label{one tree}
\end{figure}

\begin{figure}[!h]
\centering
\includegraphics[width=3.4in]{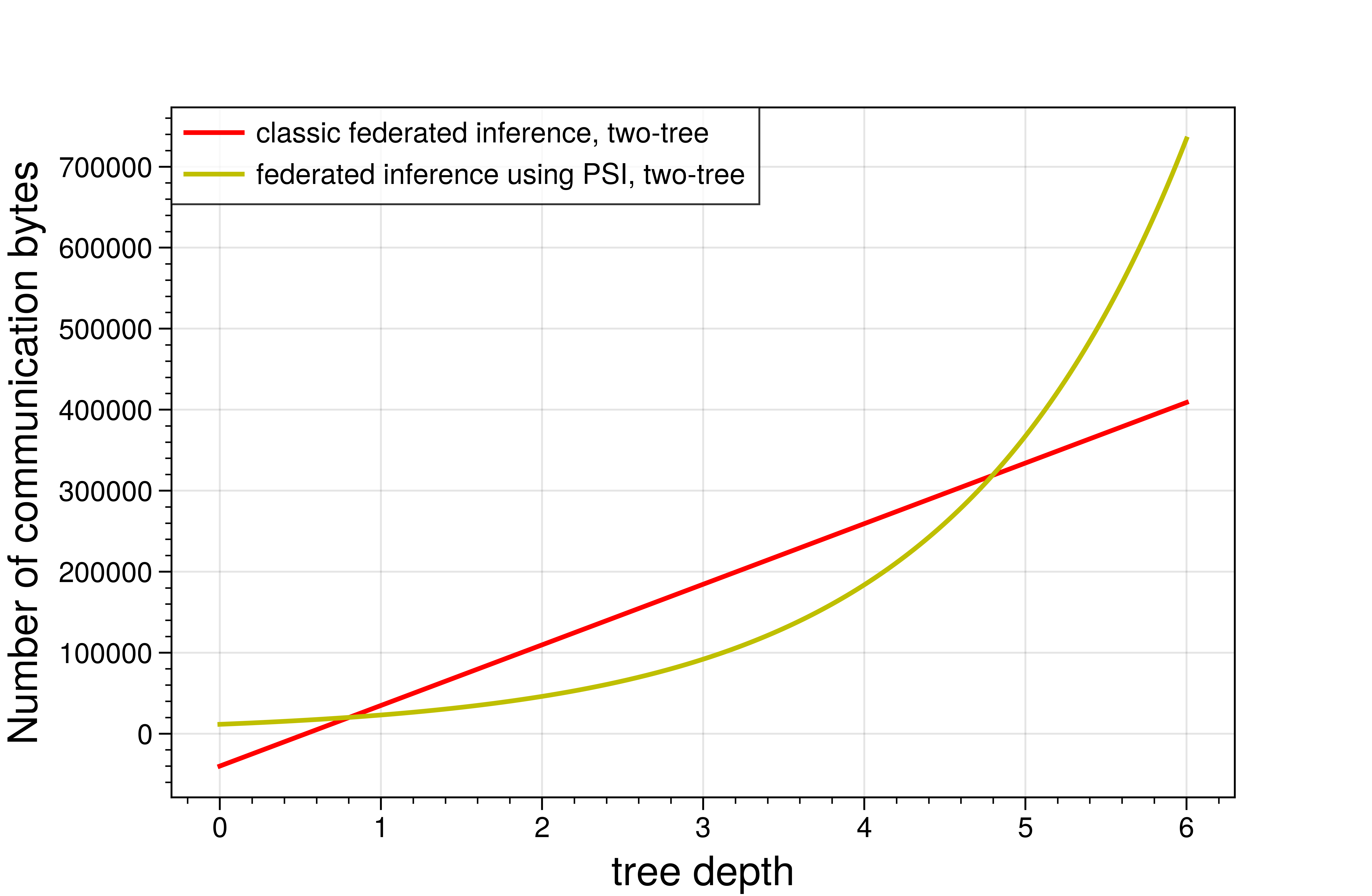}
\caption{Comparison of communications under two tree.}
\label{two tree}
\end{figure}

\begin{figure}[!h]
\centering
\includegraphics[width=3.4in]{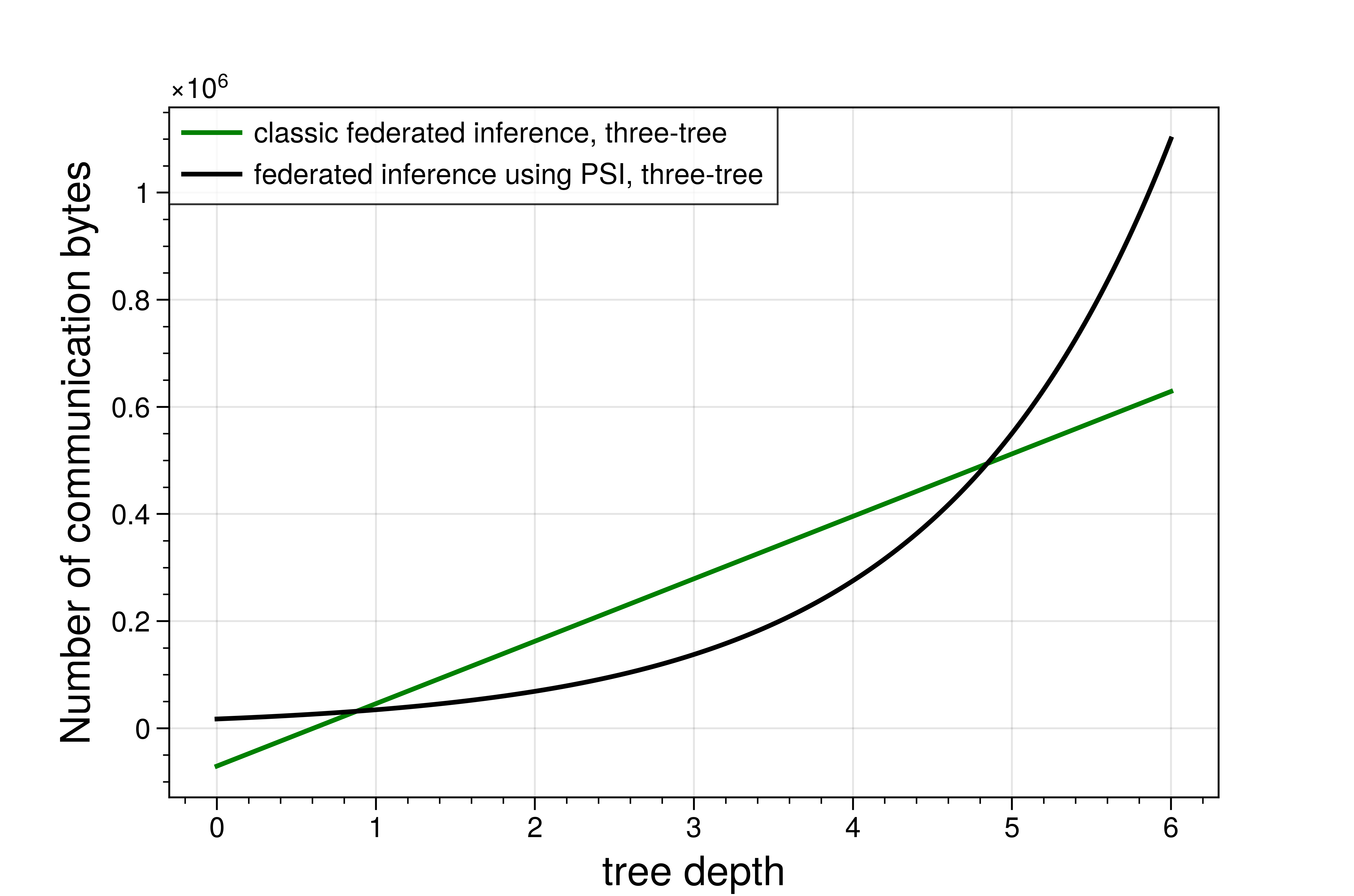}
\caption{Comparison of communications under three tree.}
\label{three tree}
\end{figure}

 \textbf{Experimental phase: }To contrast federated inference utilizing PSI against classical federated inference, we performed experiments in two scenarios: one entailed varying numbers of tree models, and the other involved uniform tree models but with varying depths. We quantified the data exchanged between participants throughout the federated inference process at different tree depths. The outcomes are illustrated in Figures \ref{one tree}, \ref{two tree}, and \ref{three tree}, each corresponding to cases with one, two, and three tree models, respectively.
\par \textbf{Result analysis: }Examination of Figures \ref{one tree}, \ref{two tree}, and \ref{three tree} reveals that, within depths 1 to 4, federated inference employing PSI consistently exhibits lower communication volume when compared to classical federated inference. The corresponding percentage breakdown is provided in Tables \ref{communication bytes}. The experimental analysis indicates that PSI-based federated inference significantly decreases communication volume in various depths: at depth 1, it achieves a reduction of approximately 24\% to 34\% compared to classical federated inference. At depth 2, the reduction ranges between 55\% to 58\%. Similarly, at depth 3, the reduction is approximately 47\% to 51\%, while at depth 4, it hovers around 27\% to 30\% when compared to classical federated inference.
\par From a practical perspective, in the realm of distributed computing, the communication infrastructure is often more challenging to enhance quickly compared to the computing environment. Therefore, schemes with lower communication requirements and greater practicality are better suited for federated programs.
\par However, communication is a critical bottleneck in federated networks which, when coupled with privacy concerns over sending raw data, necessitates that data generated on each device remain local. Indeed, federated networks frequently consist of a substantial number of devices, such as millions of smartphones, and network communication can be orders of magnitude slower than local computation, primarily due to constraints on resources like bandwidth, energy, and power. Thus, the communication efficiency of an algorithm emerges as a vital consideration for its suitability in federated networks. Algorithms requiring less communication offer distinct advantages in the context of federated networks. Therefore, our proposed federated inference method employing PSI proves more apt for federated networks than classical federated inference algorithms.


\begin{table}[]
\caption{Comparison of the communication overhead \label{communication bytes}}
\begin{tabular}{c|c|c|c|c}
\hline
\begin{tabular}[c]{@{}c@{}}Number\\ of\\ trees\end{tabular} & \begin{tabular}[c]{@{}c@{}}tree\\ depth\end{tabular} & \begin{tabular}[c]{@{}c@{}}Communication\\ bytes\\ of \cite{DBLP:journals/expert/ChengFJLCPY21}\end{tabular}  & \begin{tabular}[c]{@{}c@{}}Communication\\ bytes\\ of ours\end{tabular}  & \begin{tabular}[c]{@{}c@{}}Rate of\\ reduction in\\ communication\end{tabular} \\ \hline
                                                          & 1          & 15353  & 11607  & 24.4\%                                                                                \\ \cline{2-5} 
                                                          & 2          & 51614  & 23195  & 55.1\%                                                                          \\ \cline{2-5} 
1                                                         & 3          & 87875  & 46354  & 47.3\%                                                                          \\ \cline{2-5} 
                                                          & 4          & 124136 & 92633  & 27.4\%                                                                          \\ \cline{2-5} 
                                                          & 5          & 160397 & 185117 & ---                                                                               \\ \hline
                                                          & 1          & 34757  & 22995  & 33.8\%                                                                               \\ \cline{2-5} 
                                                          & 2          & 109581 & 45971  & 58.0\%                                                                          \\ \cline{2-5} 
2                                                         & 3          & 184406 & 91904  & 50.2\%                                                                          \\ \cline{2-5} 
                                                          & 4          & 259231 & 183732 & 29.1\%                                                                          \\ \cline{2-5} 
                                                          & 5          & 334055 & 367312 & ---                                                                               \\ \hline
                                                          & 1          & 45855  & 34446  & 24.9\%                                                                               \\ \cline{2-5} 
                                                          & 2          & 162428 & 68867  & 57.6\%                                                                          \\ \cline{2-5} 
3                                                         & 3          & 279001 & 137687 & 50.6\%                                                                          \\ \cline{2-5} 
                                                          & 4          & 395574 & 275277 & 30.4\%                                                                          \\ \cline{2-5} 
                                                          & 5          & 512147 & 550362 & ---                                                                               \\ \hline
\end{tabular}
\end{table}

\subsection{Security Analysis}
\subsubsection{Anti-Quantum Computing Attack Security}
\par
The classical secureboost model uses Paillier algorithm, while in this paper we use the CKKS algorithm. The Paillier algorithm was proposed in 1999 and the CKKS algorithm was proposed in 2017. No classical cracks have been found for these two algorithms. However, it should be noted that the security of the Paillier algorithm is based on the difficulty of determining the decisional composite residuosity assumption, while the security of the CKKS algorithm relies on the difficulty of the LWE problem. The former is threatened by Shor's algorithm in the quantum computing environment while the latter is considered to be secure against quantum computing attacks. As a result  both algorithms can provide security in a classical computing environment, while CKKS algorithm can provide protection consistently in a quantum computing environment.

\subsubsection{Anti-Side Channel Attack Security}
\par As previously mentioned, during the federated inference phase, the paths from the root node to the leaf nodes vary for different samples, leading to variations in the number of interactions between active and passive parties. Malicious participants can potentially exploit the quantity of communications between parties to conduct side channel analysis on the original federated inference algorithm. However, our proposed federated inference method employing PSI offers robust protection against such side channel analysis by necessitating only a single communication between active and passive parties in the federated inference phase.

\section{Federated Learning with Logistic Regression based on FHE}
\subsection{FHE-based Logistic Regression Model for Horizontal Federated Learning}
\subsubsection{Improved Logistic Regression model training based on the FHE} 
\hfill\par Logistic regression model is a classical model used in machine learning. In 2018, Kim et al. proposed a method based on  CKKS algorithm to train logistic regression (LR) models and designed a new encoding method to reduce the storage of encrypted datasets   \cite{DBLP:journals/iacr/KimSKLC18}. In this algorithm, ciphertext multiplications are the main costs in computations.  We optimized the process of gradient descent computation of this algorithm by reducing both the number and the depth of multiplications, which thus effectively improves  efficiency of the algorithm.
\setlength{\parskip}{0.5em}
\par (1) The Training Procedure proposed by Kim et al. \cite{DBLP:journals/iacr/KimSKLC18}
\par First, we introduce the FHE-based LR training procedure proposed by Kim et al. The application scenario is an outsourcing computation with two participants A and B providing data and computational power, respectively.
\par Participant A encrypts the dataset and the random initial weight vector $\beta^{\left( 0\right) }$, then sends them and A's public key to the data owner participant B. The dataset are locally encoded as a matrix \text{Z} of size $n × (f+1)$ and encrypted as $\text{ct}_{z}$ by using A's public key, and the weight vectors are copied n times to fill the plaintext slots, scaled by $2^{p}$, then encrypted as   $\text{ct}^{\left( 0\right)  }_{\beta }$. The plaintext matrix of the generated ciphertext is described as follows.
$$\text{ct}_{z} =\text{Enc} \left[ \begin{array}{cccc}z_{10}&z_{11}&\cdots &z_{1f}\\ z_{20}&z_{21}&\cdots &z_{2f}\\ \vdots &\vdots &\ddots &\vdots \\ z_{n0}&z_{n1}&\cdots &z_{nf}\end{array} \right]  ,$$
$$ \text{ ct}^{\left( 0\right)  }_{\beta } =\text{Enc} \left[ \begin{array}{cccc}\beta^{\left( 0\right)  }_{0} &\beta^{\left( 0\right)  }_{1} &\cdots &\beta^{\left( 0\right)  }_{f} \\ \beta^{\left( 0\right)  }_{0} &\beta^{\left( 0\right)  }_{1} &\cdots &\beta^{\left( 0\right)  }_{f} \\ \vdots &\vdots &\ddots &\vdots \\ \beta^{\left( 0\right)  }_{0} &\beta^{\left( 0\right)  }_{1} &\cdots &\beta^{\left( 0\right)  }_{f} \end{array} \right]  .$$
\par Participant B computes gradient descent based on the ciphertext weight vector $\text{ct}_{\beta}$ and the local data vector $\text{ct}_{z}$ sent by A to find an optimal model weight vector. The goal of each iteration is to update the model weight vector $\beta^{\left( t\right)} $ using the following loss gradient function.
\par
$$ \beta^{\left( t+1\right)  } \leftarrow \beta^{\left( t\right)  } +\frac{\alpha_{t} }{n} \sum^{n}_{i=1} \sigma \left( -\text{z}^{T}_{i} \beta^{\left( t\right)  } \right)  \cdot \text{z}_{i} .$$
\par
where $\alpha_{\left(t\right)}$ denotes the learning rate at the $t$-th iteration. Each iteration consists of the following eight steps.

\par Step 1: Multiply the ciphertext $\text{ct}_{z}$ and $\text{ct}^{\left( 0\right)  }_{\beta }$, rescale it by $p$ bits:
\par
$$ \text{ct}_{1} \leftarrow \text{ReScale} \left( \text{Mult} \left( \text{ct}_{z} ,\text{ct}^{\left( 0\right)  }_{\beta } \right);p\right).$$
\par
The output ciphertext contains the values $ z_{ij}\cdot \beta^{\left( t\right)  }_{j} $in its plaintext slots, i.e.,
$$ \text{ct}_{1} =\text{Enc} \left[ \begin{array}{cccc}z_{10}\beta^{\left( t\right)  }_{0} &z_{11}\beta^{\left( t\right)  }_{1} &\cdots &z_{1f}\beta^{\left( t\right)  }_{f} \\ z_{20}\beta^{\left( t\right)  }_{0} &z_{21}\beta^{\left( t\right)  }_{1} &\cdots &z_{1f}\beta^{\left( t\right)  }_{f} \\ \vdots &\vdots &\ddots &\vdots \\ z_{n0}\beta^{\left( t\right)  }_{0} &z_{n1}\beta^{\left( t\right)  }_{1} &\cdots &z_{nf}\beta^{\left( t\right)  }_{f} \end{array} \right]  .$$

\par Step 2: By adapting the incomplete column shifting operation to obtain the inner product $\text{z}^{T}_{i} \beta^{\left( t\right)  } $
$$ \text{ct}_{2} \leftarrow \text{Add} \left( \text{ct}_{1} ,\text{Rotate} \left( \text{ct}_{1} ;2^{j}\right)  \right).$$
\par $\text{for } j= 0,1, . . . , log(f+1)-1.$ In the obtained ciphertext $\text{ct}_{2}$, the inner product values $\text{z}^{T}_{i} \beta^{\left( t\right)  } $ in the first column and some “garbage” values which represented by $\star$ in the other columns. 
\[\text{ct}_{2} =\left[ \begin{matrix}\text{z}^{T}_{1} \beta^{(t)} &\star &\star &\star \\ \text{z}^{T}_{2} \beta^{\left( t\right)  } &\star &\star &\star \\ \vdots &\vdots &\ddots &\vdots \\ \text{z}^{T}_{n} \beta^{\left( t\right)  } &\star &\cdots &\star \end{matrix} \right]  .\]

\par Step 3: This step performs a constant multiplication in order to annihilate the garbage values. It can be obtained by computing the encoding polynomial $c\  \leftarrow \text{Encode} \left( C;p_{c}\right)  $ of the matrix
\[C=\left[ \begin{matrix}1&0&\cdots &0\\ 1&0&\cdots &0\\ \vdots &\vdots &\ddots &\vdots \\ 1&0&\cdots &0\end{matrix} \right], \]  
using the scaling factor of $2^{p_{c}}$ for some integer $p_{c}$. The parameter $p_{c}$ is chosen as the bit precision of plaintexts so it can be smaller than the parameter $p$.
\par Finally we multiply the polynomial $c$ to the ciphertext $\text{ct}_{2} $ and rescale it by $p_{c}$ bits:
\[\text{ct}_{3} \leftarrow \text{ReScale} \left( \text{CMult} \left( \text{ct}_{2} ;c\right)  ;p_{c}\right)  .\]
\par $\text{ct}_{2} $ is multiplied with $C$ to eliminate the garbage values in it. The resulting $\text{ct}_{3}$ encrypts the inner product values in the first column and zeros in the others:
\[\text{ct}_{3} =\text{Enc} \left[ \begin{array}{cccc}\text{z}^{T}_{1} \beta^{\left( t\right)  } &0&\cdots &0\\ \text{z}^{T}_{2} \beta^{\left( t\right)  }_{} &0&\cdots &0\\ \vdots &\vdots &\ddots &\vdots \\ \text{z}^{T}_{n} \beta^{\left( t\right)  }_{} &0&\cdots &0\end{array} \right]  .\]

\par Step 4: In this step, the first column of $\text{ct}_{3}$ is copied full of ciphertext slots by column shifting similar
to Step 2 but in the opposite direction. 
\[\text{ct}_{4} \leftarrow \text{Add} \left( \text{ct}_{3} ,\text{Rotate} \left( \text{ct}_{3} ;-2^{j}\right)  \right)  .\]
\par $\text{for } j= 0,1, . . . , log(f+1)-1.$ The output ciphertext is
\[\text{ct}_{4} =\text{Enc} \left[ \begin{array}{cccc}\text{z}^{T}_{1} \beta^{\left( t\right)  } &\text{z}^{T}_{1} \beta^{\left( t\right)  }_{} &\cdots &\text{z}^{T}_{1} \beta^{\left( t\right)  }_{} \\ \text{z}^{T}_{2} \beta^{\left( t\right)  }_{} &\text{z}^{T}_{2} \beta^{\left( t\right)  }_{} &\cdots &\text{z}^{T}_{2} \beta^{\left( t\right)  }_{} \\ \vdots &\vdots &\ddots &\vdots \\ \text{z}^{T}_{n} \beta^{\left( t\right)  }_{} &\text{z}^{T}_{n} \beta^{\left( t\right)  }_{} &\cdots &\text{z}^{T}_{n} \beta^{\left( t\right)  }_{} \end{array} \right]  .\]

\par Step 5: This step evaluates an approximating polynomial of the sigmoid function where $g(x)$ represents the approximate polynomial function.
\[\text{ct}_{5} \leftarrow g\left( \text{ct}_{4} \right)  . \]
The output ciphertext is
\[ \text{ct}_{5} =\text{Enc} \left[ \begin{array}{cccc}g(\text{z}^{T}_{1} \beta^{\left( t\right)  } )&g(\text{z}^{T}_{1} \beta^{\left( t\right)  }_{} )&\cdots &g(\text{z}^{T}_{1} \beta^{\left( t\right)  }_{} )\\ g(\text{z}^{T}_{2} \beta^{\left( t\right)  }_{} )&g(\text{z}^{T}_{2} \beta^{\left( t\right)  }_{} )&\cdots &g(\text{z}^{T}_{2} \beta^{\left( t\right)  }_{} )\\ \vdots &\vdots &\ddots &\vdots \\ g(\text{z}^{T}_{n} \beta^{\left( t\right)  }_{} )&g(\text{z}^{T}_{n} \beta^{\left( t\right)  }_{} )&\cdots &g(\text{z}^{T}_{n} \beta^{\left( t\right)  }_{} )\end{array} \right]  .\]

\par Step 6: Multiply the ciphertexts $\text{ct}_{5}$ and $\text{ct}_{z}$ and rescale the obtained result by $p$ bits as fol-
lows:
\[\text{ct}_{6} \leftarrow \text{ReScale} \left( \text{Mult} \left( \text{ct}_{5} ;\text{ct}^{}_{z} \right)  ;p\right)  .\]
The output ciphertext $\text{ct}_{6}$ is
\begin{equation*}
\text{ct}_{6}=\text{Enc}
\begin{bmatrix}g(\text{z}^{T}_{1} \beta^{\left( t\right)  } )\cdot z_{10} &\cdots &g(\text{z}^{T}_{1} \beta^{\left( t\right)  }_{} )\cdot z_{1f} \\
g(\text{z}^{T}_{2} \beta^{\left( t\right)  }_{} )\cdot z_{20}&\cdots &g(\text{z}^{T}_{2} \beta^{\left( t\right)  }_{} )\cdot z_{2f}\\ \vdots  &\ddots &\vdots \\ g(\text{z}^{T}_{n} \beta^{\left( t\right)  }_{} )\cdot z_{n0}&\cdots &g(\text{z}^{T}_{n} \beta^{\left( t\right)  }_{} )\cdot z_{nf} \end{bmatrix}.
\end{equation*}

\par Step 7: To compute the gradient of the loss function, $\text{ct}_{7}$ is obtained by recursively adding $\text{ct}_{6}$ to its row shifting:
\[\text{ct}_{7} \leftarrow \text{Add} \left( \text{ct}_{6} ,\text{Rotate} \left( \text{ct}_{6} ;2^{j}\right)  \right)  .\]
$\text{for } j= log(f + 1), . . . , log(f + 1) + log n - 1$. The output ciphertext $\text{ct}_{7}$ is
\begin{equation*}
\text{Enc}
\begin{bmatrix}\sum g(\text{z}^{T}_{i} \beta^{\left( t\right)  } )\cdot z_{i0}&\cdots &\sum g(\text{z}^{T}_{i} \beta^{\left( t\right)  }_{} )\cdot z_{if}\\ \sum g(\text{z}^{T}_{i} \beta^{\left( t\right)  }_{} )\cdot z_{i0}&\cdots &\sum g(\text{z}^{T}_{i} \beta^{\left( t\right)  }_{} )\cdot z_{if}\\ \vdots &\ddots &\vdots \\ \sum g(\text{z}^{T}_{i} \beta^{\left( t\right)  }_{} )\cdot z_{i0}&\cdots &\sum g(\text{z}^{T}_{i} \beta^{\left( t\right)  }_{} )\cdot z_{if}\end{bmatrix}  .
\end{equation*}
\par Step 8:  It uses the parameter $p_{c}$ to compute the scaled learning rate $\Delta^{\left( t\right)  } =\lfloor 2^{p_{c}}\cdot \alpha_{t} \rceil .$  The participant B updates $\beta^{\left( t\right)  } $ using the ciphertext $\text{ct}_{7}$ and the constant
$\Delta^{\left( t\right)  }$:
\[\text{ct}_{8} \leftarrow \text{ReScale} \left( \Delta^{\left( t\right)  } \cdot \text{ct}_{7} ;p_{c}\right)  ,\]
\[\text{ct}^{\left( t+1\right)  }_{\beta } \leftarrow \text{Add} \left( \text{ct}^{(t)}_{\beta } ;\text{ct}_{8} \right)  .\]
\par Finally it returns a ciphertext encrypting the updated modeling vector
\[\text{ct}^{\left( t+1\right)  }_{\beta } =\text{Enc} \left[ \begin{matrix}\beta^{\left( t+1\right)  }_{0} &\beta^{\left( t+1\right)  }_{1} &\cdots &\beta^{\left( t+1\right)  }_{f} \\ \beta^{\left( t+1\right)  }_{0} &\beta^{\left( t+1\right)  }_{1} &\cdots &\beta^{\left( t+1\right)  }_{f} \\ \vdots &\vdots &\ddots &\vdots \\ \beta^{\left( t+1\right)  }_{0} &\beta^{\left( t+1\right)  }_{1} &\cdots &\beta^{\left( t+1\right)  }_{f} \end{matrix} \right] . \]
\par where $\beta^{\left( t+1\right)  }_{j} =\beta^{\left( t\right)  }_{j} +\frac{\alpha_{t} }{n} \sum g\left( \text{z}^{T}_{i} \beta^{\left( t\right)  } \right)  \cdot z_{ij.}$
\setlength{\parskip}{0.5em}
\par (2) Summation by Rotating in CKKS
\par The above training procedure contains four multiplications separately at step 1,3,6,8. By analyzing these steps, we can find out that the multiplications at step 1 and step 6 are necessary for LR training where weights and data are multiplied together. At step 8,  the learning rate $\Delta^{\left( t\right)  }$  is multiplied to control the speed of gradient descent. While the multiplication at step 3 is brought by CKKS algorithm, because the CKKS algorithm encrypts the whole vector as one ciphertext. Thus the calculation of the inner product in CKKS can not be performed directly, but needs to be calculated by multiplication  and summation by rotating.
The study of the gradient calculation process in the paper \cite{DBLP:journals/iacr/KimSKLC18} reveals that the summation by rotating in the FHE calculation is divided into the summation of rows by rotating and the summation of columns by rotating. Different summations methods are selected according to different computational needs. The following process is uses the assumptions mentioned in the paper \cite{DBLP:journals/iacr/KimSKLC18}, which makes it possible to encrypt the entire dataset in a single ciphertext.
\par (a) Summation of Rows by Rotating
\par Let a matrix \text{Z} of size $n × (f + 1)$ be as follows:
\[\text{z} =\left[ \begin{matrix}z_{11}&z_{12}&\cdots &z_{1f}\\ z_{21}&z_{22}&\cdots &z_{2f}\\ \vdots &\vdots &\ddots &\vdots \\ z_{n1}&z_{n2}&\cdots &z_{nf}\end{matrix} \right]. \]
\par Perform the following summation by rotating operation on \text{Z} to obtain $\text{Z}_{1}$:
\par
\[\text{z}_{1} \leftarrow \text{Add} \left( \text{z} ,\text{Rotate} \left( \text{z} ;2^{j}\right)  \right). \]
\[\text{for } j= 0,1, ...,log(n)-1:\]
\[ \text{z}_{1} =\left[ \begin{matrix}\sum^{f}_{i=1} z_{1i}&\star &\star &\star \\ \sum^{f}_{i=1} z_{2i}&\star &\star &\star \\ \vdots &\vdots &\ddots &\vdots \\ \sum^{f}_{i=1} z_{ni}&\star &\cdots &\star \end{matrix} \right].  \]
\par The first column of $\text{Z}_{1}$ is the sum of the elements of each row of \text{Z}, where $\star$ stands for garbage data.
\par (b) Summation of Columns by Rotating
\par Transpose the matrix \text{Z} to obtain $\text{Z}^{T}$:
\[ \text{z}^{T} =\left[ \begin{matrix}z_{11}&z_{21}&\cdots &z_{n1}\\ z_{12}&z_{22}&\cdots &z_{n2}\\ \vdots &\vdots &\ddots &\vdots \\ z_{1f}&z_{2f}&\cdots &z_{nf}\end{matrix} \right] .\]
\par Perform the following summation by rotating operation on $\text{Z}^{T}$ to obtain $\text{Z}_{2}$:
\[ \text{z}_{2} \leftarrow \text{Add} \left( \text{z}^{T}_{} ,\text{Rotate} \left( \text{z}^{T} ;2^{j}\right)  \right) . \]
\[\text{for } j= log(f + 1), . . . , log(f + 1) + log n - 1:\]
\[ \text{z}_{2} =\left[ \begin{matrix}\sum^{f}_{i=1} z_{1i}&\sum^{f}_{i=1} z_{2i}&\cdots &\sum^{f}_{i=1} z_{ni}\\ \sum^{f}_{i=1} z_{1i}&\sum^{f}_{i=1} z_{2i}&\cdots &\sum^{f}_{i=1} z_{ni}\\ \vdots &\vdots &\ddots &\vdots \\ \sum^{f}_{i=1} z_{1i}&\sum^{f}_{i=1} z_{2i}&\cdots &\sum^{f}_{i=1} z_{ni}\end{matrix} \right].   \]
\par From the above equations, we can see that when computing summation of rows by rotating, we will get the  results and garbage data at the same time. To clear the garbage data, an extra multiplication is needed.
\par More specifically,   the authors\cite{DBLP:journals/iacr/KimSKLC18} calculate the gradient by first using a summation of rows by rotating, and then using a summation of column by rotating, after using the summation of rows by rotating, because of the presence of garbage data, it need to multiply a ciphertext constant matrix of the same dimension like $C$, whose has 1 in the first column and 0 in the rest of columns. After that, the first column of the result is copied n times to fill the ciphertext slot. After our observation and analysis, if we first transpose the weights matrix and the data matrix  at the beginning, then we can use the summation of column by rotating   and  then the summation of rows by rotating, that is, switch the order of using the two summation by rotating. In this way, we can combine multiplications in step 6 and step 8 into one calculation. So we can reduce the number of multiplications, and reduce the depth required for the whole procedure in order to improve the computational efficiency.
\setlength{\parskip}{0.5em}
\par (3) Improved Training Procedure for LR Models in FHE
\par The improved gradient descent homomorphic computation process is as follows.

\par Step 1: Transpose $\text{ct}_{z} $, transpose $\beta$ and fill the plaintext slots.
\par
$$\text{ct}_{z^{T\  }} =\left( \text{ct}_{z} \right)^{T}  =\text{Enc} \left[ \begin{array}{cccc} z_{10}&z_{20}&\cdots &z_{n0}\\ z_{11}&z_{21}&\cdots &z_{n1}\\ \vdots &\vdots &\ddots &\vdots \\ z_{1f}&z_{2f}&\cdots &z_{nf}\end{array} \right]  , $$
$$\text{ct}^{\left( 0\right)  }_{\beta^{T} } =\left( \text{ct}^{\left( 0\right)  }_{{}_{\beta }} \right)^{T}  =\text{Enc} \left[ \begin{array}{cccc}\beta^{\left( 0\right)  }_{0} &\beta^{\left( 0\right)  }_{0} &\cdots &\beta^{\left( 0\right)  }_{0} \\ \beta^{\left( 0\right)  }_{1} &\beta^{\left( 0\right)  }_{1} &\cdots &\beta^{\left( 0\right)  }_{1} \\ \vdots &\vdots &\ddots &\vdots \\ \beta^{\left( 0\right)}_{f} &\beta^{\left( 0\right)  }_{f} &\cdots &\beta^{\left( 0\right)  }_{f} \end{array} \right].$$
\par 
Step 2: Multiply the ciphertext $\text{ct}_{z^{T}}$ and $\text{ct}^{\left( 0\right) }_{\beta^{T} } $ and rescale the obtained result by $p$ bits as follows.
\par
\[\text{ct}_{1} \leftarrow \text{ReScale} \left( \text{Mult} \left( \text{ct}_{z^{T}} ;\text{ct}^{\left( t\right)  }_{\beta^{T} } \right)  ;p\right).\]

\par  Step 3: Rotate the column sum of $\text{ct}_{1} $ to get $\text{ct}_{2}, $

$$ \text{ct}_{2} \leftarrow \text{Add} \left( \text{ct}_{1} ,\text{Rotate} \left( \text{ct}_{1} ;2^{j}\right)  \right).$$

$$\text{for } j= log(f + 1), . . . , log(f + 1) + log n - 1:$$

$$\text{ct}_{2} =\text{Enc} \left[ \begin{array}{cccc}\text{z}^{T}_{1} \beta^{\left( t\right)  } &\text{z}^{T}_{2} \beta^{\left( t\right)  }_{} &\cdots &\text{z}^{T}_{n} \beta^{\left( t\right)  }_{} \\ \text{z}^{T}_{1} \beta^{\left( t\right)  }_{} &\text{z}^{T}_{2} \beta^{\left( t\right)  }_{} &\cdots &\text{z}^{T}_{n} \beta^{\left( t\right)  }_{} \\ \vdots &\vdots &\ddots &\vdots \\ \text{z}^{T}_{1} \beta^{\left( t\right)  }_{} &\text{z}^{T}_{2} \beta^{\left( t\right)  }_{} &\cdots &\text{z}^{T}_{n} \beta^{\left( t\right)  }_{} \end{array} \right] .$$ 
\par 
Step 4: Calculate the approximate polynomial of the sigmoid function, where g(x) is the approximate function.
\[\text{ct}_{3} \leftarrow g\left( \text{ct}_{2} \right)  . \]

\par 
Step 5: Multiply the ciphertexts $\text{ct}_{z^{T}}$ and $\text{ct}_{3}$ and rescale the obtained result by p bits as follows.
$$ \text{ct}_{4} \leftarrow \text{ReScale} \left( \text{Mult} \left( \text{ct}_{3} ;\text{ct}^{}_{z^{T}} \right)  ;p\right).$$

\par 
Step 6: Rotate the row sum of $\text{ct}_{4}$ to get $\text{ct}_{5}$.
$$ \text{ct}_{5} \leftarrow \text{Add} \left( \text{ct}_{4} ,\text{Rotate} \left( \text{ct}_{4} ;2^{j}\right)  \right).$$

$$\text{for } j= 0,1, . . . , log(f+1)-1:$$

$$\text{ct}_{5} =\text{Enc} \left[ \begin{array}{cccc}\sum^{}_{} g(\text{z}^{T}_{i} \beta^{\left( t\right)  } )\cdot z_{i0}&\star &\cdots &\star \\ \sum^{}_{} g(\text{z}^{T}_{i} \beta^{\left( t\right)  }_{} )\cdot z_{i1}&\star &\cdots &\star \\ \vdots &\vdots &\ddots &\vdots \\ \sum^{}_{} g(\text{z}^{T}_{i} \beta^{\left( t\right)  }_{} )\cdot z_{if}&\star &\cdots &\star \end{array} \right].$$
\par 
Step 7: This step removes the garbage values while multiplying the learning rate. Where the learning rate: $\Delta^{\left( t\right)  } =\lfloor 2^{p_{c}}\cdot \alpha_{t} \rceil.$

\[\text{ct}_{6} \leftarrow \text{ReScale} \left( \text{Mult} \left( \text{c} ;\text{ct}^{}_{5} \right)  ;p_{c}\right)  . \]
\[ \text{c} =\left[ \begin{matrix}2^{p_{c}}\cdot \alpha_{t} &0&\cdots &0\\ 2^{p_{c}}\cdot \alpha_{t} &0&\cdots &0\\ \vdots &\vdots &\ddots &\vdots \\ 2^{p_{c}}\cdot \alpha_{t} &0&\cdots &0\end{matrix} \right]   ,\]
$$ \text{ct}_{6} =\text{Enc} \left[ \begin{matrix}2^{p_{c}}\cdot \alpha_{t} \sum^{}_{} g(\text{z}^{T}_{i} \beta^{\left( t\right)  } )\cdot z_{i0}&0&\cdots &0\\ 2^{p_{c}}\cdot \alpha_{t} \sum^{}_{} g(\text{z}^{T}_{i} \beta^{\left( t\right)  }_{} )\cdot z_{i1}&0&\cdots &0\\ \vdots &\vdots &\ddots &\vdots \\ 2^{p_{c}}\cdot \alpha_{t} \sum^{}_{} g(\text{z}^{T}_{i} \beta^{\left( t\right)  }_{} )\cdot z_{if}&0&\cdots &0\end{matrix} \right]. $$
\par 
Step 8: Copy the first column element in $\text{ct}_{6}$ by column to fill the ciphertext slot and get $\text{ct}_{7}$, add $\text{ct}_{7}$ to $\text{ct}^{\left( t\right) }_{\beta^{T} } $ to get the final gradient: 
\[ \text{ct}_{7} \leftarrow \text{Add} \left( \text{ct}_{6} ,\text{Rotate} \left( \text{ct}_{6} ;-2^{j}\right)  \right)   ,\]
\[\text{for } j= 0,1, ...,log(n)-1:\]
\[ \text{ct}^{\left( t+1\right)  }_{\beta } \leftarrow \text{Add} \left( \text{ct}^{\left( t\right)  }_{\beta } ,\text{ct}_{7} \right)  . \]
\[ \text{ct}^{\left( t+1\right)  }_{\beta^{T} } =\text{Enc} \left[ \begin{matrix}\beta^{\left( t+1\right)  }_{0} &\beta^{\left( t+1\right)  }_{0} &\cdots &\beta^{\left( t+1\right)  }_{0} \\ \beta^{\left( t+1\right)  }_{1} &\beta^{\left( t+1\right)  }_{1} &\cdots &\beta^{\left( t+1\right)  }_{1} \\ \vdots &\vdots &\ddots &\vdots \\ \beta^{\left( t+1\right)  }_{f} &\beta^{\left( t+1\right)  }_{f} &\cdots &\beta^{\left( t+1\right)  }_{f} \end{matrix} \right].  \]
\par  After calculating the updated gradient from the above steps, participant B sends the resulting ciphertext to participant A, who decrypts it to update the gradient. In the training process, the main computational functions are ciphertext multiplication, ciphertext addition and vector rotation, where the computational overhead of one ciphertext multiplication is much larger than one ciphertext addition or vector rotation. Therefore, our optimization scheme can effectively increase   the training efficiency by reducing the number of ciphertext multiplication at the cost of two extra transpositions which have very low overhead because they are calculated in plain text form.  And the comparison results are shown in Table \ref{LRimprove}.
\begin{table}[h!]
\centering
\caption{Comparisons of LR training calculation\label{LRimprove}}
%\resizebox{\textwidth}{!}{
\begin{tabular}{c|c}
\hline
 & \cite{DBLP:journals/iacr/KimSKLC18}     \\
\hline\hline
num of mul& 4   \\
\hline
num of add &$\log_{2} \left( n\times (f+1)\right)  +\log_{2} n+3$ \\
\hline
num of rotation & $\log_{2} \left( n\times (f+1)\right)  +\log_{2} n+2$ \\
\hline
depth of mul& 5  \\
\hline
& Ours \\
\hline\hline
num of mul & 3\\
\hline
num of add & $\log_{2} \left( n\times (f+1)\right)+\log_{2} n+3$\\
\hline
num of rotation & $\log_{2} \left( n\times (f+1)\right)  +\log_{2} n+2$\\
\hline
depth of mul & 4 \\
\hline
\end{tabular}
%}
\end{table}

\subsubsection{FHE-based Logistic Regression Model for Horizontal Federated Learning}
\hfill\par The classical HFL used as a control in this chapter is the logistic regression-based HFL scenario from Chapter 15 of "Federated Learning" \cite{DBLP:series/synthesis/2019YangLCKCY} by Professor Qiang Yang et al.
\par In more detail, the training process of the HFL model implemented with classical HFL is as follows.
Denote  clients as A, who provide samples and train  model locally, and B represents the server, who owns the model and generates encryption key pair. In step 1, B generates Paillier key pair, sends the public key as well as encrypted model $\Theta$ to A. In step 2, A train the encrypted model using local data by utilizing  homomorphic encryption algorithm and then send the result gradient back to B. In step 3, B decrypts the gradient with the private key and performs aggregation optimization, updates the model parameters and sends them to A, which continues the model training. The above steps are repeated several times until the model training is completed when the loss reaches the expected or the number of training sessions reaches the set maximum. During the training process, each participant does not know the data structure of the other participants and can only get the parameters needed for its own part of the sample.  The training process is shown in Table \ref{classHorFL} and the evaluation process is shown in the Table \ref{classhorFLtest}.

Our design of HFL algorithm based on FHE can be obtained by replacing Paillier  by CKKS algorithm  and using the  proposed steps in section \uppercase\expandafter{\romannumeral4}.A.1) to compute LR gradient.

\begin{table}[!h]
\centering
\caption{Training process of HFL in FATE}
\label{classHorFL}
\begin{tabular}{p{20pt}|p{130pt}|p{70pt}}
\hline
 & Client A  & Server B \\
\hline\hline
Step1 &Divide the dataset into $k$ equal parts and simulate a scenario where $k$ participants train together and each participant trains the model locally individually; &Generate a pair of secret keys $P_{k_{B}},S_{k_{B}}$, and initialize $\Theta$, send $P_{k_{B}}$, $\left[ \left[ \Theta \right]  \right]  $ to A; \\
\hline
Step2 & Each participant randomly draws \text{batch-size} samples $x_{i}$ locally, compute $\left[ \left[ \frac{\partial L}{\partial \theta } \right]  \right]  =\frac{1}{n} \sum^{n}_{i=1} \left( \frac{1}{4} \left[ \left[ \theta^{T} \right]  \right]  x_{i}+\frac{1}{2} [\left[ -1]\right]  y_{i}\right) x_{i}$, send $\left[ \left[ \frac{\partial L}{\partial \Theta } \right]  \right]$ to B; & \\
\hline
Step3 &  &Decrypt $\left[ \left[ \frac{\partial L}{\partial \Theta } \right]  \right]$, perform model aggregation, update $\Theta$. \\
\hline
\end{tabular}
\end{table}

\par 
\begin{table}[!h]
\centering
\caption{Evaluation process of HFL in FATE}
\label{classhorFLtest}
\begin{tabular}{p{22pt}|p{50pt}|p{140pt}}
\hline
 & Client A & Server B \\
\hline\hline
Step1 &Send $\left[ \left[ \frac{\partial L}{\partial \Theta } \right]  \right]$ to B. & Extract the batchsize data $x_{i}$ from the test set in order, decrypt $[[\Theta]]$, compute the inner product \text{wx} of $x_{i}$ and $\Theta$;\\ 
\hline
Step2 && Compute $pred_{y_{i}}$ using the sigmoid function, The number of $pred_{y_{i}}>0.5$ is the number of correct values; \\
\hline
Step3 & & Compute the accuracy $acc$ by  Number of correct / Total number of evaluation. \\
\hline
\end{tabular}
\end{table}

\subsubsection{Security Analysis}
%The security of the scheme is described next.
\hfill\par (1) Anti-Quantum Computing Attack Security
\par As can be seen from the algorithm flow, in the HFL algorithm based on FHE, the input and output forms of the two participants are the same as the classical HFL, i.e., the model side sends the model parameters in ciphertext, and the data side obtains the gradient and sends it to the model side by means of ciphertext calculation of the model parameters and local data. The difference lies in the encryption algorithm of the data and the ciphertext training process, so if we assume that the security of the fully homomorphic CKKS algorithm is not lower than that of the partially homomomorphic Paillier algorithm, we can obtain that the security of the HFL algorithm based on the CKKS algorithm is not lower than that of the classical HFL based on the Paillier algorithm under semi-honest model.
\par
The Paillier algorithm was proposed in 1999 and the CKKS algorithm was proposed in 2017. No classical cracks have been found for these two algorithms. However, it should be noted that the security of the Paillier algorithm is based on the difficulty of determining the decisional composite residuosity assumption, while the security of the CKKS algorithm relies on the difficulty of the LWE problem. The former is threatened by Shor's algorithm in the quantum computing environment while the latter is considered to be secure against quantum computing attacks. As a result  both algorithms can provide security in a classical computing environment, while   CKKS algorithm can provide protection consistently in a quantum computing environment.
\par
\par (2) Anti-Gradient Attack Security
\par
In the classical federated learning algorithm, the data owner avoids data leaving the local area by sending gradients to prevent data leakage, but related works \cite{DBLP:journals/corr/abs-2001-02610, 2020Analyzing, DBLP:conf/nips/GeipingBD020, 2018Exploiting, 2017Deep} show  that gradients can leak data information, and sending gradients alone cannot protect data security. To solve this problem, using the gradient perturbation method in differential privacy to resist gradient gradient attacks is a feasible approach.
\par Unlike other FHE algorithms  which provides exact results,   CKKS algorithm perform approximate computations. This is not deviating from the demand, because most of the operations in practical problems, dealing with real numbers (complex numbers), which often only need to retain a part of the valid numbers. In addition, allowing for errors and relaxing the accuracy limits makes CKKS a greater simplification of details and a significant improvement in computational efficiency compared to other homomorphic schemes based on the LWE/RLWE problem. Since the approximate computational characteristics of CKKS that the data is encrypted with a Gaussian noise perturbation of controllable size is added synchronously. The size of the perturbation affects the accuracy of the model, but on the other hand it also provides a layer of security for the data.

\subsubsection{Performance Analysis}
\hfill\par (1) Application \uppercase\expandafter{\romannumeral1}: Cancer prediction (medical field) 
\par \textbf {Dataset: }This experiment uses the breast cancer dataset \cite{Federated-Learning}, which contains 569 samples with 30 features. The data samples are derived from measurements of the breast lump image and whether it is cancerous or not, and the goal is to use these measurements to predict whether the lump is cancerous or not. In this dataset, the number of positive and negative samples is balanced, so the model quality can be assessed using accuracy.
\par \textbf{Experimental environment: }Intel(R) Core(TM) i9-10980XE CPU @ 3.00GHz
\par
\begin{figure}[!h]
\centering
\includegraphics[width=3.4in]{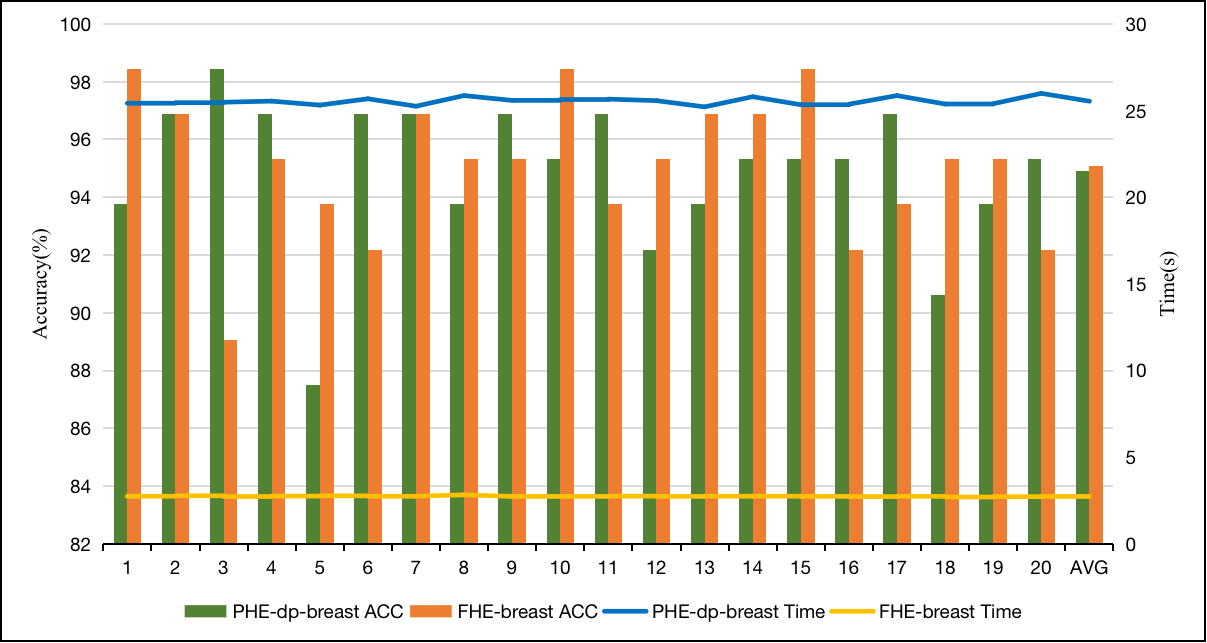}
\caption{Comparison of efficiency and accuracy in breast cancer dataset.}
\label{fig_1}
\end{figure}

 \textbf{Experimental phase: }The control experiments use the classical HFL code based on the Paillier algorithm implemented in  "Federated Learning in Practice" by Yang et al\cite{Federated-Learning} with differential privacy strategy propsed in \cite{DBLP:journals/corr/abs-2001-02610}. We compared the experimental results under the same parameters, including model accuracy and model training time (excluding communication time), for one round of local training of the model. The experimental results of the classical federated learning algorithm and our algorithm are shown in Figure \ref{fig_1}.

\par \textbf{Results Analysis: }As can be seen in Figure \ref{fig_1}, the average training time for the model based on PHE is 25.5s, and the average training time for the model based on FHE is 2.7s. the training efficiency of the FHE-based HFL algorithm is 9.3 times higher than that of the classical HFL algorithm, and the average accuracy of our scheme is slightly improved.

\subsection{FHE-based Logistic Regression Model for Vertical Federated Learning}
\subsubsection{Dataset pre-processing}
\hfill\par (1) Privacy Set Intersection
\par Private set intersection (PSI)\cite{2004Efficient} refers to the task of jointly computing the set intersection between two parties holding their own sets of project data, without revealing any information beyond the intersection. Each participants of VFL have independent data, so they usually need to filter the data with the same id by PSI before performing model training.
\par According to the recent results, the PSI protocol based on multipoint inadvertent pseudo-random function construction proposed by Chase et al.\cite{DBLP:conf/crypto/ChaseM20} is the current practical solution with excellent computational performance and communication overhead, so we implement our algorithm's PSI protocol based on it.

\par (2) Binning and Calculating WOE Values based on the FHE
\par Data binning is a common method of data pre-processing in machine learning, also known as discrete binning or data segmentation. Its essence is to group the data according to specific rules to achieve data discretization, enhance data stability and reduce the risk of overfitting. Logistic regression is a generalized linear model with limited expressiveness. The main reason why accuracy of logistic regression models can be effectively improved by data binning is that after data binning, the univariate is discretized into N variables and each variable has a separate weight, which is equivalent to introducing a nonlinear module for the model and can improve the model expression and fitting ability, so data binning is one of the common methods in the pre-processing stage of logistic regression models.

\par For continuous features, data binning is mainly divided into unsupervised binning and supervised binning. The unsupervised binning mainly includes isometric binning and isofrequency binning, and the supervised binning mainly includes decision tree binning and cardinality binning. After experimental analysis isometric binning is more suitable for the application scenario of this paper, so the following experiments use isometric binning.

\par WOE (Weight of Evidence) is the logarithm of the proportion of positive and negative samples under a certain value for a character-based variable or a segment for a continuous variable. It is a form of coding for the original independent variables. To encode a variable with WOE, it is necessary to first bin the variable. The formula for calculating the WOE value is as follows.
\[ WOE\  =\  \text{ln} \left( \frac{p_{y_{1}}}{p_{y_{0}}} \right)  =\text{ln} \left( \frac{\frac{G_{i}}{G_{T}} }{\frac{B_{i}}{B_{T}} } \right). \]
Where $G_{i}$ denotes the number of positive samples in this one bin and $G_{T}$ denotes the total number of positive samples in all bins. $B_{i}$ denotes the number of negative samples in this one bin and $B_{T}$ denotes the total number of negative samples in all bins.
\par As seen above, in the process of calculating the WOE values, it is necessary to obtain the number of positive and negative samples in each bin of a feature. For HFL, the participants have their respective labels and can directly bin and compute WOE values under plaintext, but for VFL, only one participant  has the labels and cannot compute WOE values directly.   With the help of FHE, we propose a novel method   for data binning and computing WOE values and the details are shown below.
\par Denote  the labels as y, $x_i$ is the features of a column without labels obtained by binning, and $x_{bin} $ is its corresponding bin matrix.
\begin{equation}\label{eq1}
 \begin{matrix}y&&&&x_{i}\\ 1&&&&1\\ 0&&&&2\\ 1&&&&2\\ 1&&&&3\\ 0&&&&1\end{matrix} \begin{matrix}&\\ &\\ &\\ &\\ &\\ &\end{matrix} \begin{matrix}x_{bin}\\ \left[ \begin{matrix}1&0&0\\ 0&1&0\\ 0&1&0\\ 0&0&1\\ 1&0&0\end{matrix} \right]  \end{matrix}
 \end{equation}

We can compute WOE values through following steps  in Table \ref{He_woe}.
\begin{table}[!h]
\centering
\caption{The process of computing WOE values based on FHE for dataset binning}
\label{He_woe}
\begin{tabular}{p{22pt}|p{100pt}|p{95pt}}
\hline
 & Party A  & Party B \\
\hline\hline
Step 1 &
Read the dataset $x_{A}$, bin and compute the WOE value under plaintext;
&Generate a pair of keys $P_{k_{B}},S_{k_{B}}$ and read the dataset $x_{B}$ to bin the dataset. After the binning, the bin matrix is generated according to the formula (\ref{eq1}) and the bin matrix is encrypted in chunks as $[[Chunk\_bins]]$. Send the public keys $P_{k_{B}}$, $[[Chunk\_bins]]$ to A;  \\
\hline
Step 2 & Fill the plaintext slots with y values and encrypt $[[y]]$, the number of positive samples per bin in B $[[Good\_num_i]]= \sum([[y_i]]*[[Chunk\_bins_i]])$. Send $[[Good\_num_i]]$ to B. &The total number of samples in each bin $Bin\_total\_num_i$ can be obtained by summing the columns of the bin matrix. Decrypt $[[Good\_num_i]]$, $Bad\_num_i=(Bin\_total\_num_i) - (Good\_num_i)$, and just calculate the WOE value by the formula.\\
\hline
\end{tabular}
\end{table}

\par (3) SMOTE Algorithm based on FHE
\par SMOTE algorithm\cite{2011SMOTE} is a sample derivation algorithm that expands the dataset based on the relationship between samples. It can expand minority samples in the dataset according to   existing samples to achieve a balanced number of positive and negative samples, thus avoiding the model bias caused by the unbalanced samples. For example, in a financial fraud transaction detection scenario, the order samples of fraudulent transactions usually represent a very small fraction of the total number of transactions, but identifying these few samples is crucial to the detection task, so the SMOTE algorithm can be used to expand the percentage of minority samples.
\par From the principle of SMOTE algorithm, first we need to know whether a sample is a minority sample, that is, we need to know the label value $y$ in order to expand the sample. For HFL, the sample expansion can be achieved by directly calling the SMOTE function in plain text, but for VFL, only one party has the label information and cannot expand the sample directly. Based on this, we propose a way to calculate SMOTE function based on FHE. The details of calculation procedure is shown in Table \ref{He_smote} (in this case, we assume the positive samples are minority without losing the generality).

\begin{table}[!h]
\centering
\caption{SMOTE algorithm based on FHE}
\label{He_smote}
\begin{tabular}{p{22pt}|p{120pt}|p{70pt}}
\hline
 & Party A  & Party B \\
\hline\hline
Step 1 & &Fill with 0 for data position in A, encrypt $x_{B}$ in chunks. Send $[[x_{B}]]$ to A;  \\
\hline
Step 2 & Fill with 0 for data position in B, encrypt $x_{A}$ in chunks. Get all positive samples $samples_A$ in A and the positive sample index good\_num\_index. compute $[[x_{i}]]=[[x_{A}]]+[[x_{B}]]$;& \\
\hline
Step 3 & Use $samples_A$ to find the $k$ nearest neighbor index $neig\_index$ of each positive sample, and copy $k$ copies of each index of $samples_A$ to get $orig\_index$, which corresponds to $neig\_index$ one by one. The corresponding values are extracted from $[[x_{i}]$ by multiplication and Rotation to $[[orig\_sample]]$ and $[[neig\_sample]]$ respectively according to the two indexes;& \\
\hline
Step 4 & Generate the same random number matrix $\Lambda$ for each row and compute $[[new\_sample]]=\Lambda[[neig\_sample]]+(Mat_1-\Lambda)[[orig\_sample]],$ where $Mat_1$ is the matrix with all values of $1$. Generate a random matrix $R=\{R_A||R_B\}$, where $R_A,R_B$ are the same size as the matrix of both samples, compute $[[new\_sample\_R]]=[[new\_sample]]+[[R]]$, send $[[new\_sample\_R]]$ to B;& \\
\hline
Step 5 & & Decrypt $[[new\_sample\_R]]$, divide $new\_sample\_R$ into $A\_sample\_R$ and $B\_sample$ according to the number of AB features, send $A\_sample\_R$ to A. Add $B\_sample$ to the dataset $x_{B}$.\\
\hline
Step 6 & Compute $A\_sample = A\_sample\_R - R_A$, add $A\_sample$ to the dataset $x_{A}$, and save $R_B$ for training.& \\
\hline
\end{tabular}
\end{table}

\subsubsection{FHE-based Logistic Regression Model for Vertical Federated Learning}
\setlength{\parskip}{0.5em}
\hfill\par (1) Classical Vertical Federated Learning Logistic Regression Model
\par To compare with classical federated learning, we first introduce the VFL in the FATE framework. FATE (Federated AI Technology Enabler) is the world's first industrial-grade open source framework for federated learning initiated by WeBank, which enables enterprises and organizations to collaborate on data while protecting data security and data privacy. The FATE project uses Multiparty Secure Computing (MPC) and Homomorphic Encryption (HE) technologies to build an underlying secure computing protocol that supports secure computing for different kinds of machine learning, including logistic regression, tree-based algorithms, deep learning, and migration learning. 
\par The classical logistic regression-based VFL model training process is as follows.
\par A represents the initiator, B represents the participant, and C, as the third party coordinating the work, is responsible for generating the private and public keys. The encryption algorithm chooses Paillier homomorphic encryption, and the C neutral party generates the secret key pair and sends the public key to A and B. The private key is in the C party. In the information interaction, A and B transmit the information encrypted with the public key to C. C decrypts the information with the private key and transmits the aggregated optimized gradient to A and B respectively, and A, B update the model parameters and continue the model training. The above steps are repeated until the model training is completed when the loss reaches the expected or the training times reach the set maximum. 
\par In the logistic regression algorithm, let $u^{A}_{i}=w_{A}x_{A},u^{B}_{i}=w_{B}x_{B}$, loss is:
$$loss\approx log2-\frac 12yw^Tx+\frac 18\left(w^Tx\right)^2,$$ the residual term d(the coefficient in the gradient calculation):
$$d=\left(\frac {1}{1+exp(-yw^Tx)}-1\right)y\approx\left(\frac 12yw^Tx-1\right)\frac 12y,$$
gradient $g$ is:
$$g=\left(\frac {1}{1+exp(-yw^Tx)}-1\right)yx\approx\left(\frac 12yw^Tx-1\right)\frac 12yx,$$ where $w^Tx=\left(w_A x_A+w_B x_B\right),\left(w^Tx\right)^2=\left(w_A x_A\right)^2+\left(w_B x_B\right)^2+2w_A x_A w_B x_B$
, local encryption gradient $g_A,g_B$ is:
$$g_A=\left[\left[\frac{\partial L}{\partial \Theta^A}\right]\right]=\left[\left[d\right]\right]*x^A,g_B=\left[\left[\frac{\partial L}{\partial \Theta^B}\right]\right]=\left[\left[d\right]\right]*x^B,$$ 
as shown in table \ref{FATEtraining}.
\begin{table}[!h]
\centering
\caption{VFL training process in FATE}
\label{FATEtraining}
\begin{tabular}{p{22pt}|p{60pt}|p{60pt}|p{60pt}}
\hline
 & Party A  & Party B & Party C \\
\hline\hline
Step 1 & Compute $\left[\left[u_A\right]\right] and \left[\left[u_A^2\right]\right]$, send them to B; & &\\
\hline
Step 2 & & Compute $d$ and loss, send $\left[\left[d\right]\right]$ to A; & \\
\hline
Step 3 & Compute local encryption gradient $\left[\left[g_A\right]\right]$, send $\left[\left[g_A\right]\right]$ to C; & Compute gradient $\left[\left[g_B\right]\right]=X^{H^j}\left[\left[d\right]\right]$, send $\left[\left[g_B\right]\right]$ and $\left[\left[loss\right]\right]$ to C; &\\
\hline
Step 4 & & & Decrypt $\left[\left[loss\right]\right]$ and end the model training if it converges. Decrypt $g_A$ and $g_B$ send them back to parties A and B.\\
\hline
Step 5 & update $\theta_A$. & update $\theta_B$. & \\
\hline
\end{tabular}
\end{table}
\par Implementation of logistic regression evaluation process in VFL with FATE: Investigating the evaluation code of both ABs, it can be obtained that the evaluation method in this framework needs to be carried out jointly with the model evaluation data of both parties. The participating party sends its own data for $w^{T}x$ calculation to the initiating party in plaintext form, and the initiating party then adds its own feature results to make predictions, and the process is shown in Table \ref{FATEevaluating}.
\begin{table}[!h]
\centering
\caption{FATE VFL logistic regression evaluation process}
\label{FATEevaluating}
\begin{tabular}{p{22pt}|p{60pt}|p{60pt}|p{60pt}}
\hline
 & Party A  & Party B & Party C \\
\hline\hline
Step 1 & & Compute $u_B$, send it to A. & \\
\hline
Step 2 & Prediction with $u_A$ and $u_B$. & & \\
\hline
\end{tabular}
\end{table}

\par (2) Vertical Federated Learning Logistic Regression Model based on FHE
\par The analysis of the classical VFL process shows that the main reason why the algorithm requires the participation of a third party is that the PHE algorithm used to compute the gradient only supports the computation of ciphertext-plaintext multiplication and requires the assistance of a trusted third party to compute the gradient.

\begin{figure}[!h]
\centering
\includegraphics[width=3.4in]{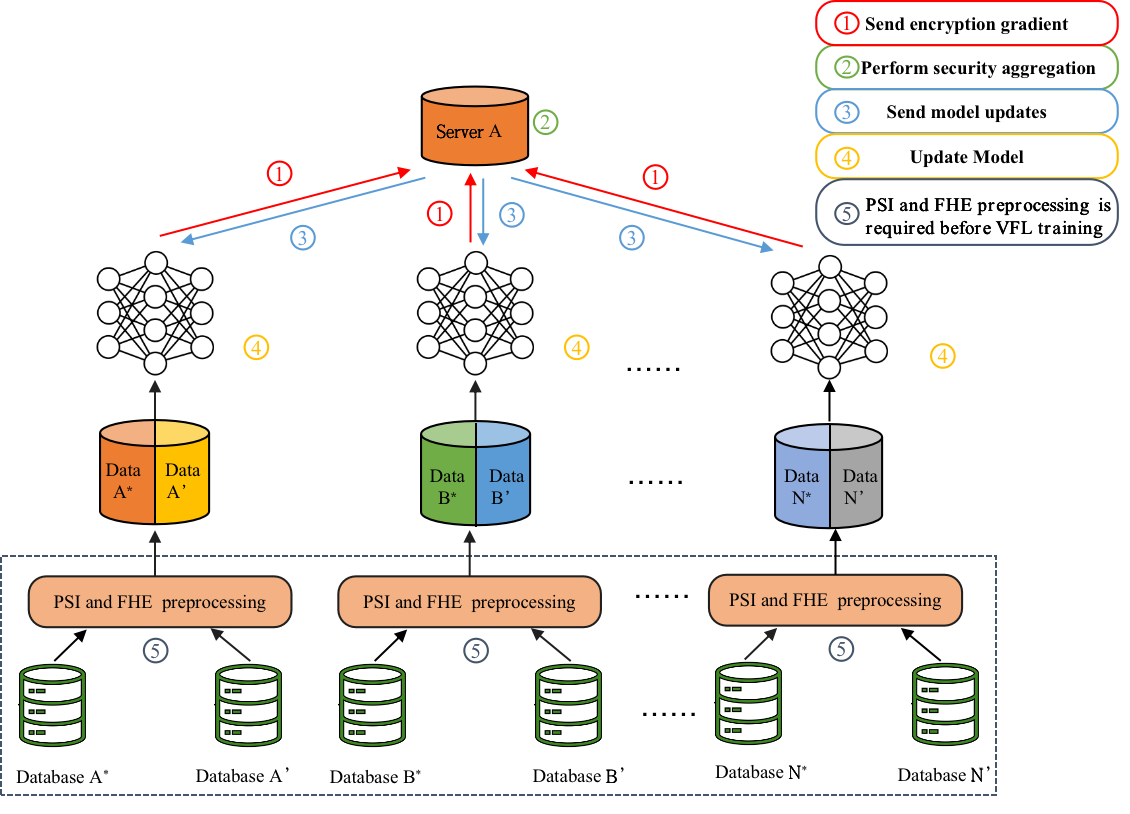}
\caption{Unified federated learning framework.}
\label{v_H_fhe}
\end{figure}

When using the FHE algorithm, it can accomplish the VFL training goal without a trusted third party by FHE's property that can realize the ciphertext-ciphertext multiplication computation, and this property can significantly improve the current situation that VFL is difficult to implement due to the lack of a trusted third party, and effectively improve the practicality. In addition, we can design a unified horizontal/vertical federated learning framework based on FHE accordingly, so that horizontal/vertical federated learning can be invoked under the same framework with a high degree of reuse, and there is no need to deploy separate versions, the unified framework is shown in Figure \ref{v_H_fhe}. For example, VFL based on FHE can accomplish the following training objectives.
\par Suppose Company A and Company B want to jointly train a machine learning model, and they both have their own data in their business systems. In addition, Company A has labels for the data needed to predict the model. Using B as a server for model training, the training process is divided into the following 4 steps:
\par(a) A and B find the data intersection by PSI and send the encrypted data to A;
\par(b) A computes training gradients locally, masks the selection of gradients using encryption, differential privacy or secret sharing techniques, and sends the masking results to the server;
\par(c) B performs security aggregation without knowing any information about A;
\par(d) B decrypts the gradient and updates the model;

\par Likewise, we use logistic regression and homomorphic encryption as examples to illustrate the training process. In order to train a logistic regression model with gradient descent, we need to safely compute its loss and gradient.
 Suppose the learning rate $\eta$, the regularization parameter $\lambda$, the data set $\{x^{A}_{i},y_{i}\}_{i\in D_{A}} ,\{ x^{B}_{i}\}_{i\in
 D_{B}}$, the model parameters are $\Theta$, corresponding to the feature space corresponding to $x_{i}$ obtained by PSI. The loss function is as follows.
 \par
 \[\left[ \left[ L\right]  \right]  =\frac{1}{n} \sum^{n}_{i=1} \{ \log 2-\frac{1}{2} y_{i}[[\theta^{T} ]]([[x_{A}]]\]
 \[+\left[ \left[ x_{B}\right]  \right]  )+\frac{1}{8} \left( [[\theta^{T} ]]([[x_{A}]]+\left[ \left[ x_{B}\right]  \right]  )\right)^{2}  \} .     \]
\par The intersection of $ x^{A}_{i}$ and $ x^{B}_{i}$ is obtained by PSI, for A, the feature values belonging only to B are complemented by 0, and for B, the feature values belonging only to A are complemented by 0. The result is spelled into as $ x_{i}$. We use about 80\% of the data in $x_{i}$ for training and about 20\%  for testing.
The gradient descent formula based on FHE is as follows:
\[ \left[ \left[ \frac{\partial L}{\partial \theta } \right]  \right]  =\frac{1}{n} \sum^{n}_{i=1} \left( \frac{1}{4} \left[ \left[ \theta^{T} \right]  \right]  [[x_{i}]]+\frac{1}{2} [\left[ -1]\right] y_{i}\right)  [[x_{i}]].   \]
\par
The training process is shown in Table \ref{OurFL}.
\par

\begin{table}[!h]
\centering
\caption{VFL training process based on FHE}
\label{OurFL}
\begin{tabular}{p{22pt}|p{100pt}|p{90pt}}
\hline
 & Party A  & Party B \\
\hline\hline
Step 1 & &Generate a pair of secret keys $P_{k_{B}},S_{k_{B}}$, initialize $\Theta$, extract \text{batch-size} samples from the B dataset according to the random sequence generated by the random seed, and encrypt them into $[[x_{B}]]$ after preprocessing such as filling with 0 and transposition.
Send the public keys $P_{k_{B}}$, $\left[ \left[ \Theta \right] \right] $ and $[[x_{B}]]$ to A; \\
\hline
Step 2 & Extract \text{batch-size} samples $x_{A}$ from the A dataset according to the random sequence generated by the random seed, encrypting $x_{A}$ after preprocessing and adding $[[x_{B}]]$ to $[[x_{A}]]$ to get $[[x_{i}]]$, compute $\left[ \left[ \frac{\partial L}{\partial \theta } \right]  \right] =\frac{1}{n} \sum^{n}_{i=1} $$\left( \frac{1}{4} \left[ \left[ \theta^{T} \right]  \right]  [[x_{i}] +\frac{1}{2} [\left[ -1]\right]  y_{i}\right) [[x_{i}]]$.& \\
\hline
Step 3 &  &Decrypt $\left[ \left[ \frac{\partial L}{\partial \Theta } \right]  \right]$, update $\Theta$.\\
\hline
\end{tabular}
\end{table}

\par In the evaluation process, we also adopted a new approach to evaluate in order to prevent privacy leakage. The detail process is shown in Table \ref{OurFLtest}.
\begin{table}[!h]
\centering
\caption{VFL evaluation process based on FHE}
\label{OurFLtest}
\begin{tabular}{p{22pt}|p{100pt}|p{90pt}}
\hline
 & Party A  & Party B \\
\hline\hline
Step 1 &  Compute the inner product [[wx]] of $[[x_{i}]]$ and $[[\Theta]]$;& \\ 
\hline
Step 2 & Using the first-order Taylor approximation sigmoid function, we obtain: $[[pred_{y_{i}}]]=0.5+0.25[[wx]]$. Set the function$[[f\left( pred_{y_{i}}, y_{i}\right) ]]=\left( [[pred_{y_{i}}]]-0.5\right)  y_{i}$, then $[[f\left( pred_{y_{i}}, y_{i}\right)]]=0.25[[wx]]\cdot y_{i}$;&  \\
\hline
Step 3 & Randomly generate $R_{i}\in \left( 0,1\right) $ , let $[[pred_{R_{i}}]]=[[f]]\cdot R_{i}$, send $[[pred_{R_{i}}]]$ to B. &  Decrypt to get $pred_{R_{i}}$. The number of $pred_{R_{i}}>0$ is the number of correct values;\\
\hline
Step 4 & & Calculate the accuracy acc by Number of correct/Total number of evaluation.\\
\hline
\end{tabular}
\end{table}

\subsubsection{Security Analysis}
\hfill\par (1) Anti-Quantum Computing Attack and Anti-gradient Attack Security
\par From the above flow of the VFL algorithm based on FHE, we can see that we unify the HEL and VFL models by using the characteristics of FHE, so that the training process of the two is basically the same, and the difference mainly lies in whether the model side needs to provide the additional amount of features it has. Therefore, HFL based on FHE can be regarded as a special case of vfl based on FHE, so that we are able to base the security of both algorithms on the same security assumptions, and therefore our VFL scheme is as secure anti-quantum computing attacks and anti-gradient attacks as HFL. This unified process structure can effectively reduce security risks and provide easier deployment conditions.

\par (2) Security without Trusted Third-Party Assumptions
%{\color{red}这段不就是简单描述吗
\par When there is a third party C in federated learning, C assigns the same public key to both sides of the computation(data), A and B, at the very beginning, and the third party C holds the corresponding private key.
Party A and Party B make corresponding calculations on their own data, encrypt the intermediate results based on their own data with public keys, and send the ciphertext to the third party C.
C uses the private key to decrypt the ciphertext of intermediate results from each party to obtain the plaintext of intermediate results from each party, and aggregates them to get the complete intermediate results. This process is iterated until the end condition is satisfied. From the final result, the third party C obtains information that it should not have obtained during the participation process, and this information may expose the private data of other participants.
\par In contrast to schemes with third-party participation, our scheme does not rely on the trusted third-party assumption. The VFL scheme based on FHE theoretically requires that the participating parties do not have access to additional information, except for their own data and each other's data after homomorphic encryption. Only the information between the participants interacts, and the interaction process during the scheme execution does not leak the private data of the participants.

\par (3) Model Evaluation of Security based on FHE Algorithm
\par In VFL, each party holds part of the features of the sample, so that no participant can do the model accuracy evaluation independently. Implementing the logistic regression evaluation process in classical VFL requires the participants to compute $w\dot x$ and send it to the initiator in plaintext form. Since $w\dot x$ contains part of the information of $x$, $x$ can be fully recovered when specific conditions are met, e.g., when $b=w'x$ is computed multiple times using different $w'$, a system of equations can be built and the full information of $x$ can be determined by Gaussian elimination method, so this evaluation method has obvious security risks. In contrast, the model evaluation algorithm based on FHE is executed by ciphertext computation, and its security will not be reduced by repeated training, which can effectively guarantee data security.

\subsubsection{Performance Analysis} 

\hfill\par (1) Application \uppercase\expandafter{\romannumeral2}: Voice recognition (biometric field)
\par \textbf{Dataset: }This dataset \cite{gender-voice} was created to identify whether the voice is male or female based upon the acoustic properties of the voice and speech. The dataset consists of 3168 recorded speech samples, each possessing 20 features, collected from male and female speakers. 
\par \textbf{Preprocessing: }We split the sound recognition dataset into two parties A, B, where each sample in A contains label values and 13 features, and each sample in B contains the remaining 7 features. And the number of samples in A and B are expanded to 4800 respectively by filling random samples before performing PSI. Since the positive and negative examples in the samples are balanced, the accuracy can be used to judge the model usability.
\par \textbf{Experimental environment: }Intel(R) Core(TM) i9-10980XE CPU @ 3.00GHz
\par
\begin{figure}[!h]
\centering
\includegraphics[width=3.4in]{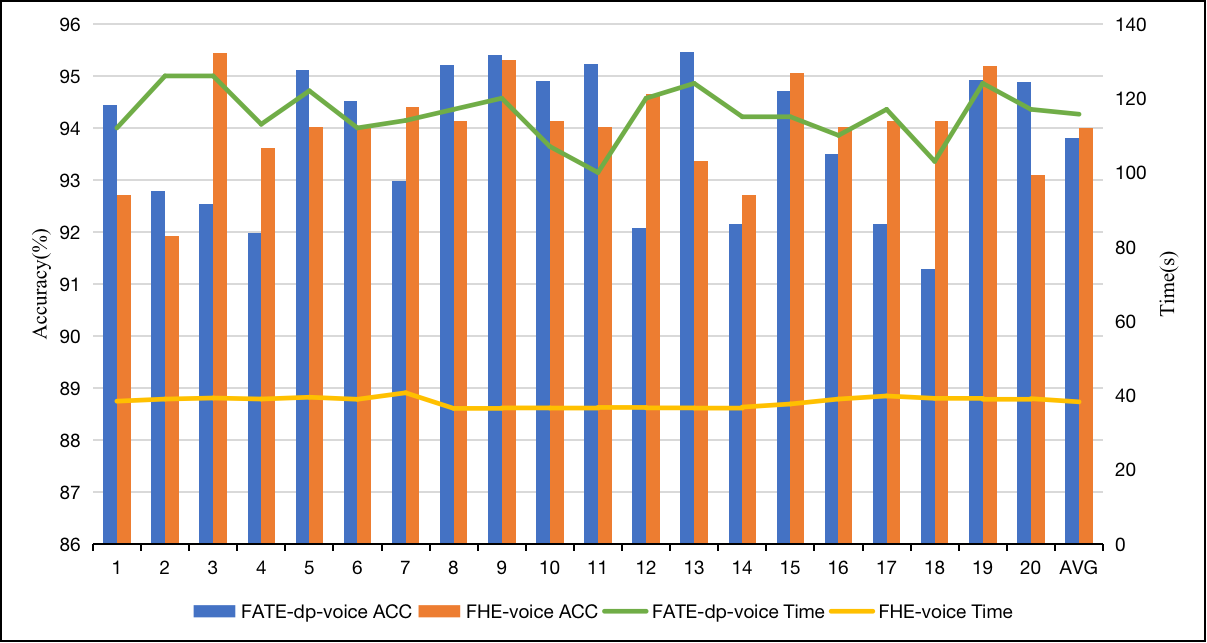}
\caption{Comparison of efficiency and accuracy in voice recognition dataset.}
\label{fig_2}
\end{figure}

 \textbf{Experimental phase: }Under the fully homomorphic VFL based on logistic regression, we were conducted 20 experiments, the model training time (excluding communication time consumption) and the accuracy of the model are obtained respectively. The control group used the FATE version 1.6.0 VFL scheme with differential privacy strategy propsed in \cite{DBLP:journals/corr/abs-2001-02610}, and the results obtained under the same dataset are shown in Figure \ref{fig_2}.
\par \textbf{Result analysis: }By analyzing the experimental data, we can see that the average training time of the VFL scheme based on FHE is 38.2s, and the average accuracy is 94\%; while the average training time of the classical VFL scheme with the same dataset is 115.7s, and the average accuracy is 93.8\%. Thus, the training efficiency of our scheme is 3 times higher than that of classical VFL in this dataset and the accuracy also improved slightly, with an average range of 0.2\%.

%\section{Vertical Federate Algorithm based on FHE}
%\subsection{VFL module based on FHE}
\hfill\par (2) Application \uppercase\expandafter{\romannumeral3}: Company bankruptcy prediction (financial field)
\par \textbf{Dataset: }The data \cite{company-bankruptcy-prediction}were collected from the Taiwan Economic Journal for the years 1999 to 2009. Company bankruptcy was defined based on the business regulations of the Taiwan Stock Exchange. The data includes 6819 samples, each containing 95 features.
\par \textbf{Preprocessing: }We split the company bankruptcy dataset into two parties A, B. Each sample in A contains label values and 60 features, and each sample in B contains the remaining 35 features. Firstly, we bin and calculate the WOE value in FHE for the dataset. We experimentally conclude that isometric binning performs better on this dataset, and the more bins can get higher accuracy, so we use isometric binning for the experiment. Since the positive samples in this dataset only account for 3.2\% of all samples, we use the SMOTE function based on FHE to expand the dataset to 13731 samples by adding positive samples. To simulate the real situation where there are other samples in addition to the intersection of A and B, the number of samples in A and B are expanded to 17,000 respectively by filling random samples before PSI, and the accuracy can be used to judge the usability of the model due to the balance of positive and negative cases in the samples. In particular, since the FATE algorithm only supports the WOE function but not the SMOTE function, and the FATE-based control group cannot train a usable model for this dataset using only the WOE function, we used the same WOE and SMOTE pre-processed dataset to conduct a comparison experiment for model training.
\par \textbf{Experimental environment: }Intel(R) Core(TM) i9-10980XE CPU @ 3.00GHz
\par
\begin{figure}[!h]
\centering
\includegraphics[width=3.4in]{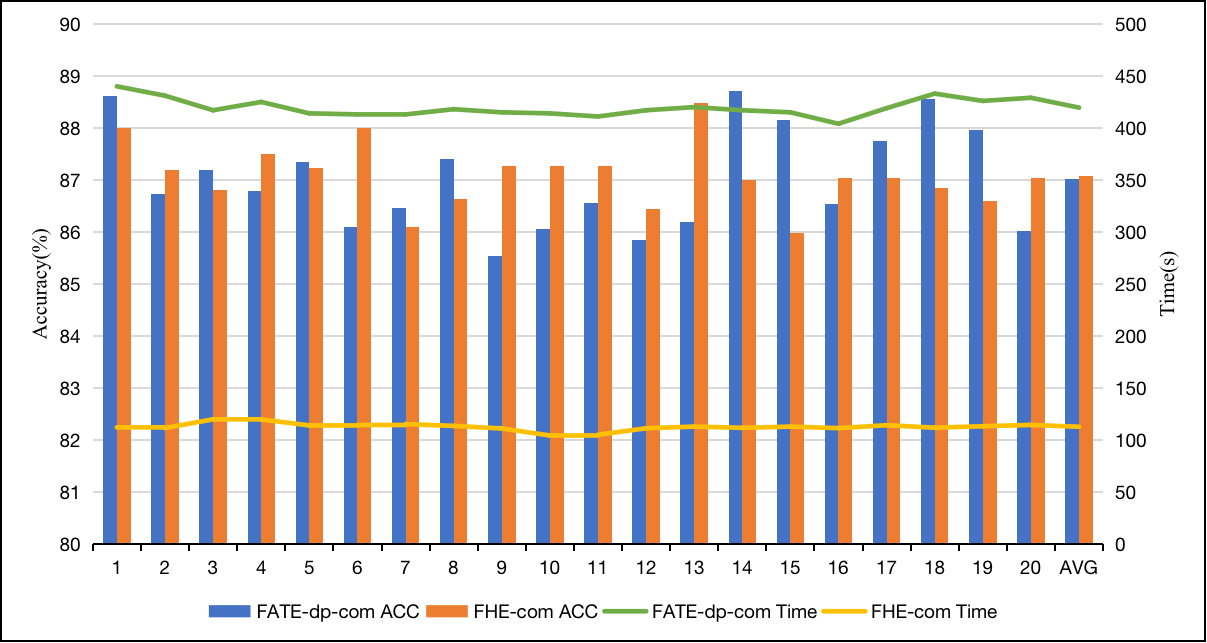}
\caption{Comparison of efficiency and accuracy in company bankruptcy dataset.}
\label{fig_3}
\end{figure}

 \textbf{Experimental phase: }Under the fully homomorphic VFL based on logistic regression, we were conducted 20 experiments, the model training time (excluding communication time consumption) and the accuracy of the model are obtained respectively. The control group used the FATE version 1.6.0 VFL scheme with differential privacy strategy propsed in \cite{DBLP:journals/corr/abs-2001-02610}, and the results obtained under the same dataset are shown in Figure \ref{fig_3}.
\par \textbf{Result analysis: }By analyzing the experimental data, we can see that the average training time of the VFL scheme based on FHE is 112.6s, and the average accuracy is about 87.1\%; while the average training time of the classical VFL scheme with the same dataset is 419.6s, and the average accuracy is about 87\%. Thus, the training efficiency of our scheme is 3.7 times higher than that of classical VFL in this dataset and the accuracy also improved slightly, with an average range of 0.1\%.

%\section{Vertical Federate Algorithm based on FHE}
%\subsection{VFL module based on FHE}

\section{Conclusion}
\setlength{\parskip}{0.5em}
In this paper we combine fully homomorphic encryption and federated learning techniques and propose a set of horizontal and vertical federated learning schemes based on fully homomorphic encryption, which achieve greater advantages over classical federated learning in terms of functionality, security and efficiency:

\par \textbf{More Functions:} Our federated learning schemes support both tree model and linear model. At the same time, based on the functionality of the FHE algorithm, our federated learning schemes can support high precision approximation of complex loss functions and can cope with more complex training objectives and tasks. Our federated learning schemes provide protection of models and data for a wider range of application scenarios. Our SecureBoost model for vertical federated learning provides a more secure and less communicative approach to federated inference, and our Logistic Regression model for vertical federated learning can support model evaluation operations initiated by any participant. In addition, our proposed federated learning scheme based on FHE achieves unification in the training process, and participants can perform horizontal/vertical federated learning operations according to data distribution in the same framework. Therefore, there is no need to deploy separate horizontal/vertical versions. We also implement the computation of WOE values and the SMOTE algorithm based on FHE, allowing our federated learning to be applied in scenarios where the dataset is highly unbalanced between positive and negative samples, which is not possible with classical federated learning. 
\par \textbf{Better Security:} The security of our federated learning schemes rely on the lattice problem, which is resistant to quantum computing attacks and is the security assumption adopted by the post-quantum public key standard published by the National Institute of Standards and Technology (NIST), so that our schemes can provide protection in future quantum computing environments. In addition, CKKS algorithm provides controlled error protection for the data, so our federated learning schemes are as effective as the classical federated learning schemes with differential privacy protection in terms of security against gradient attacks, and can better protect the data of each participant.
\par \textbf{More efficient:} With a well-designed algorithmic process and precise parameter selection, our proposed federated learning schemes obtain significant improvements in training efficiency over the classical federated learning schemes. Experimental results show that our secureboost federated learning model is 1.4-2 times more efficient than the classical algorithm, the horizontal logistic regression federated learning model is 9.3 times more efficient than the classical algorithm for training, and the vertical logistic regression federated learning model is 3-3.7 times more efficient than the classical algorithm for training.

\bibliography{IEEE-Transactions-LaTeX2e-templates-and-instructions/ref}

\begin{thebibliography}{10}
\providecommand{\url}[1]{#1}
\csname url@samestyle\endcsname
\providecommand{\newblock}{\relax}
\providecommand{\bibinfo}[2]{#2}
\providecommand{\BIBentrySTDinterwordspacing}{\spaceskip=0pt\relax}
\providecommand{\BIBentryALTinterwordstretchfactor}{4}
\providecommand{\BIBentryALTinterwordspacing}{\spaceskip=\fontdimen2\font plus
\BIBentryALTinterwordstretchfactor\fontdimen3\font minus \fontdimen4\font\relax}
\providecommand{\BIBforeignlanguage}[2]{{%
\expandafter\ifx\csname l@#1\endcsname\relax
\typeout{** WARNING: IEEEtran.bst: No hyphenation pattern has been}%
\typeout{** loaded for the language `#1'. Using the pattern for}%
\typeout{** the default language instead.}%
\else
\language=\csname l@#1\endcsname
\fi
#2}}
\providecommand{\BIBdecl}{\relax}
\BIBdecl

\bibitem{DBLP:conf/aistats/McMahanMRHA17}
B.~McMahan, E.~Moore, D.~Ramage, S.~Hampson, and B.~A. y~Arcas, ``Communication-efficient learning of deep networks from decentralized data,'' in \emph{Proceedings of the 20th International Conference on Artificial Intelligence and Statistics, {AISTATS} 2017, 20-22 April 2017, Fort Lauderdale, FL, {USA}}, ser. Proceedings of Machine Learning Research, vol.~54, 2017, pp. 1273--1282.

\bibitem{MEMBER1981On}
MEMBER, IEEE, D.~Dolev, and A.~C. Yao, ``On the security of public key protocols,'' \emph{Information Theory IEEE Transactions on}, vol.~29, no.~2, pp. 198--208, 1981.

\bibitem{10.1145/1866307.1866390}
\BIBentryALTinterwordspacing
W.~Dai, H.~Jin, D.~Zou, S.~Xu, W.~Zheng, and L.~Shi, ``Tee: A virtual drtm based execution environment for secure cloud-end computing,'' in \emph{Proceedings of the 17th ACM Conference on Computer and Communications Security}, ser. CCS '10.\hskip 1em plus 0.5em minus 0.4em\relax New York, NY, USA: Association for Computing Machinery, 2010, p. 663–665. [Online]. Available: \url{https://doi.org/10.1145/1866307.1866390}
\BIBentrySTDinterwordspacing

\bibitem{dwork2006calibrating}
C.~Dwork, F.~McSherry, K.~Nissim, and A.~Smith, ``Calibrating noise to sensitivity in private data analysis,'' in \emph{Third Theory of Cryptography Conference (TCC 2006)}, ser. Lecture Notes in Computer Science, vol. 3876.\hskip 1em plus 0.5em minus 0.4em\relax Springer, March 2006, pp. 265--284.

\bibitem{RonaldRivest}
R.~L. Rivest, L.~Adleman, and M.~L. Dertouzos, ``On data banks and privacy homomorphisms,'' \emph{Foundations of secure computation}, vol. 4(Nov.):169-180, 1978.

\bibitem{DBLP:journals/expert/ChengFJLCPY21}
K.~Cheng, T.~Fan, Y.~Jin, Y.~Liu, T.~Chen, D.~Papadopoulos, and Q.~Yang, ``Secureboost: {A} lossless federated learning framework,'' \emph{{IEEE} Intell. Syst.}, vol.~36, no.~6, pp. 87--98, 2021.

\bibitem{olr}
F.~E. Harrell, ``Ordinal logistic regression,'' vol. In Regression modeling strategies.\hskip 1em plus 0.5em minus 0.4em\relax Springer, 2001, p. 331–343.

\bibitem{2021Advances}
E.~Kairouz and H.~B. Mcmahan, ``Advances and open problems in federated learning,'' \emph{Foundations and Trends® in Machine Learning}, vol.~14, no.~1, 2021.

\bibitem{Li_2023}
Q.~Li, Z.~Wen, Z.~Wu, S.~Hu, N.~Wang, Y.~Li, X.~Liu, and B.~He, ``A survey on federated learning systems: Vision, hype and reality for data privacy and protection,'' \emph{{IEEE} Transactions on Knowledge and Data Engineering}, vol.~35, no.~4, pp. 3347--3366, apr 2023.

\bibitem{XIA2021100008}
``A survey of federated learning for edge computing: Research problems and solutions,'' \emph{High-Confidence Computing}, vol.~1, no.~1, p. 100008, 2021.

\bibitem{DBLP:journals/ftml/KairouzMABBBBCC21}
P.~Kairouz, H.~B. McMahan, B.~Avent, A.~Bellet, M.~Bennis, A.~N. Bhagoji, K.~A. Bonawitz, Z.~Charles, G.~Cormode, R.~Cummings, R.~G.~L. D'Oliveira, and H.~Eichner, ``Advances and open problems in federated learning,'' \emph{Found. Trends Mach. Learn.}, vol.~14, no. 1-2, pp. 1--210, 2021.

\bibitem{DBLP:journals/tii/FengLYGW22}
C.~Feng, B.~Liu, K.~Yu, S.~K. Goudos, and S.~Wan, ``Blockchain-empowered decentralized horizontal federated learning for 5g-enabled uavs,'' \emph{{IEEE} Trans. Ind. Informatics}, vol.~18, no.~5, pp. 3582--3592, 2022.

\bibitem{DBLP:journals/tvcg/WangCXWZS23}
X.~Wang, W.~Chen, J.~Xia, Z.~Wen, R.~Zhu, and T.~Schreck, ``Hetvis: {A} visual analysis approach for identifying data heterogeneity in horizontal federated learning,'' \emph{{IEEE} Trans. Vis. Comput. Graph.}, vol.~29, no.~1, pp. 310--319, 2023.

\bibitem{DBLP:journals/corr/abs-1711-10677}
S.~Hardy, W.~Henecka, H.~Ivey{-}Law, R.~Nock, G.~Patrini, G.~Smith, and B.~Thorne, ``Private federated learning on vertically partitioned data via entity resolution and additively homomorphic encryption,'' \emph{CoRR}, vol. abs/1711.10677, 2017.

\bibitem{2020Privacy}
G.~Xu, H.~Li, Y.~Zhang, S.~Xu, J.~Ning, and R.~Deng, ``Privacy-preserving federated deep learning with irregular users,'' \emph{IEEE Transactions on Dependable and Secure Computing}, vol.~PP, no.~99, pp. 1--1, 2020.

\bibitem{DBLP:conf/eurocrypt/Paillier99}
P.~Paillier, ``Public-key cryptosystems based on composite degree residuosity classes,'' in \emph{Advances in Cryptology - {EUROCRYPT} '99, International Conference on the Theory and Application of Cryptographic Techniques, Prague, Czech Republic, May 2-6, 1999, Proceeding}, ser. Lecture Notes in Computer Science, J.~Stern, Ed., vol. 1592.\hskip 1em plus 0.5em minus 0.4em\relax Springer, 1999, pp. 223--238.

\bibitem{STOC:gen09}
C.~Gentry, ``Fully homomorphic encryption using ideal lattices,'' in \emph{Proceedings of the forty-first annual ACM symposium on Theory of computing}, 2009, pp. 169--178.

\bibitem{DBLP:conf/crypto/Brakerski12}
Z.~Brakerski, ``Fully homomorphic encryption without modulus switching from classical gapsvp,'' in \emph{Advances in Cryptology - {CRYPTO} 2012 - 32nd Annual Cryptology Conference, Santa Barbara, CA, USA, August 19-23, 2012. Proceedings}, ser. Lecture Notes in Computer Science, vol. 7417.\hskip 1em plus 0.5em minus 0.4em\relax Springer, 2012, pp. 868--886.

\bibitem{brakerski2014efficient}
Z.~Brakerski and V.~Vaikuntanathan, ``Efficient fully homomorphic encryption from (standard) lwe,'' \emph{SIAM Journal on computing}, vol.~43, no.~2, pp. 831--871, 2014.

\bibitem{ITCS:BraGenVai12}
Z.~Brakerski, C.~Gentry, and V.~Vaikuntanathan, ``(leveled) fully homomorphic encryption without bootstrapping,'' in \emph{Proceedings of the 3rd Innovations in Theoretical Computer Science Conference -- ITCS'12}.\hskip 1em plus 0.5em minus 0.4em\relax ACM, 2012, p. 309–325.

\bibitem{DBLP:conf/crypto/GentrySW13}
C.~Gentry, A.~Sahai, and B.~Waters, ``Homomorphic encryption from learning with errors: Conceptually-simpler, asymptotically-faster, attribute-based,'' in \emph{Advances in Cryptology - {CRYPTO} 2013 - 33rd Annual Cryptology Conference, Santa Barbara, CA, USA, August 18-22, 2013. Proceedings, Part {I}}, ser. Lecture Notes in Computer Science, vol. 8042.\hskip 1em plus 0.5em minus 0.4em\relax Springer, 2013, pp. 75--92.

\bibitem{DBLP:conf/asiacrypt/CheonKKS17}
J.~H. Cheon, A.~Kim, M.~Kim, and Y.~S. Song, ``Homomorphic encryption for arithmetic of approximate numbers,'' in \emph{Advances in Cryptology - {ASIACRYPT} 2017 - 23rd International Conference on the Theory and Applications of Cryptology and Information Security, Hong Kong, China, December 3-7, 2017, Proceedings, Part {I}}, ser. Lecture Notes in Computer Science, T.~Takagi and T.~Peyrin, Eds., vol. 10624.\hskip 1em plus 0.5em minus 0.4em\relax Springer, 2017, pp. 409--437.

\bibitem{DBLP:journals/iacr/DemmlerRRT18}
\BIBentryALTinterwordspacing
D.~Demmler, P.~Rindal, M.~Rosulek, and N.~Trieu, ``{PIR-PSI:} scaling private contact discovery,'' \emph{{IACR} Cryptol. ePrint Arch.}, p. 579, 2018. [Online]. Available: \url{https://eprint.iacr.org/2018/579}
\BIBentrySTDinterwordspacing

\bibitem{DBLP:journals/fgcs/LvYYCFLLLLZ20}
\BIBentryALTinterwordspacing
S.~Lv, J.~Ye, S.~Yin, X.~Cheng, C.~Feng, X.~Liu, R.~Li, Z.~Li, Z.~Liu, and L.~Zhou, ``Unbalanced private set intersection cardinality protocol with low communication cost,'' \emph{Future Gener. Comput. Syst.}, vol. 102, pp. 1054--1061, 2020. [Online]. Available: \url{https://doi.org/10.1016/j.future.2019.09.022}
\BIBentrySTDinterwordspacing

\bibitem{DBLP:conf/bibm/Shen0WFD18}
\BIBentryALTinterwordspacing
L.~Shen, X.~Chen, D.~Wang, B.~Fang, and Y.~Dong, ``Efficient and private set intersection of human genomes,'' in \emph{{IEEE} International Conference on Bioinformatics and Biomedicine, {BIBM} 2018, Madrid, Spain, December 3-6, 2018}, H.~J. Zheng, Z.~Callejas, D.~Griol, H.~Wang, X.~Hu, H.~H. H.~W. Schmidt, J.~Baumbach, J.~Dickerson, and L.~Zhang, Eds.\hskip 1em plus 0.5em minus 0.4em\relax {IEEE} Computer Society, 2018, pp. 761--764. [Online]. Available: \url{https://doi.ieeecomputersociety.org/10.1109/BIBM.2018.8621291}
\BIBentrySTDinterwordspacing

\bibitem{DBLP:conf/sp/Meadows86}
C.~A. Meadows, ``A more efficient cryptographic matchmaking protocol for use in the absence of a continuously available third party,'' in \emph{Proceedings of the 1986 {IEEE} Symposium on Security and Privacy, Oakland, California, USA, April 7-9, 1986}.\hskip 1em plus 0.5em minus 0.4em\relax {IEEE} Computer Society, 1986, pp. 134--137.

\bibitem{DBLP:conf/ccs/AranhaLO022}
\BIBentryALTinterwordspacing
D.~F. Aranha, C.~Lin, C.~Orlandi, and M.~Simkin, ``Laconic private set-intersection from pairings,'' in \emph{Proceedings of the 2022 {ACM} {SIGSAC} Conference on Computer and Communications Security, {CCS} 2022, Los Angeles, CA, USA, November 7-11, 2022}, H.~Yin, A.~Stavrou, C.~Cremers, and E.~Shi, Eds.\hskip 1em plus 0.5em minus 0.4em\relax {ACM}, 2022, pp. 111--124. [Online]. Available: \url{https://doi.org/10.1145/3548606.3560642}
\BIBentrySTDinterwordspacing

\bibitem{DBLP:conf/ccs/DongCW13}
\BIBentryALTinterwordspacing
C.~Dong, L.~Chen, and Z.~Wen, ``When private set intersection meets big data: an efficient and scalable protocol,'' in \emph{2013 {ACM} {SIGSAC} Conference on Computer and Communications Security, CCS'13, Berlin, Germany, November 4-8, 2013}, A.~Sadeghi, V.~D. Gligor, and M.~Yung, Eds.\hskip 1em plus 0.5em minus 0.4em\relax {ACM}, 2013, pp. 789--800. [Online]. Available: \url{https://doi.org/10.1145/2508859.2516701}
\BIBentrySTDinterwordspacing

\bibitem{DBLP:conf/crypto/ChaseM20}
M.~Chase and P.~Miao, ``Private set intersection in the internet setting from lightweight oblivious {PRF},'' in \emph{Advances in Cryptology - {CRYPTO} 2020 - 40th Annual International Cryptology Conference, {CRYPTO} 2020, Santa Barbara, CA, USA, August 17-21, 2020, Proceedings, Part {III}}, ser. Lecture Notes in Computer Science, D.~Micciancio and T.~Ristenpart, Eds., vol. 12172.\hskip 1em plus 0.5em minus 0.4em\relax Springer, 2020, pp. 34--63.

\bibitem{2016XGBoost}
T.~Chen and C.~Guestrin, ``Xgboost: A scalable tree boosting system,'' \emph{ACM}, 2016.

\bibitem{Federated-Learning}
https://github.com/FederatedAI/Practicing Federated-Learning.

\bibitem{misc_wholesale_customers_292}
M.~Cardoso, ``{Wholesale customers},'' UCI Machine Learning Repository, 2014, {DOI}: https://doi.org/10.24432/C5030X.

\bibitem{DBLP:journals/iacr/KimSKLC18}
A.~Kim, Y.~Song, M.~Kim, K.~Lee, and J.~H. Cheon, ``Logistic regression model training based on the approximate homomorphic encryption,'' \emph{{IACR} Cryptol. ePrint Arch.}, p. 254, 2018.

\bibitem{DBLP:series/synthesis/2019YangLCKCY}
Q.~Yang, Y.~Liu, Y.~Cheng, Y.~Kang, T.~Chen, and H.~Yu, \emph{Federated Learning}, ser. Synthesis Lectures on Artificial Intelligence and Machine Learning.\hskip 1em plus 0.5em minus 0.4em\relax Morgan {\&} Claypool Publishers, 2019.

\bibitem{DBLP:journals/corr/abs-2001-02610}
\BIBentryALTinterwordspacing
B.~Zhao, K.~R. Mopuri, and H.~Bilen, ``idlg: Improved deep leakage from gradients,'' \emph{CoRR}, vol. abs/2001.02610, 2020. [Online]. Available: \url{http://arxiv.org/abs/2001.02610}
\BIBentrySTDinterwordspacing

\bibitem{2020Analyzing}
M.~Song, Z.~Wang, Z.~Zhang, Y.~Song, and H.~Qi, ``Analyzing user-level privacy attack against federated learning,'' \emph{IEEE Journal on Selected Areas in Communications}, vol.~PP, no.~99, pp. 1--1, 2020.

\bibitem{DBLP:conf/nips/GeipingBD020}
J.~Geiping, H.~Bauermeister, H.~Dr{\"{o}}ge, and M.~Moeller, ``Inverting gradients - how easy is it to break privacy in federated learning?'' in \emph{Advances in Neural Information Processing Systems 33: Annual Conference on Neural Information Processing Systems 2020, NeurIPS 2020, December 6-12, 2020, virtual}, 2020.

\bibitem{2018Exploiting}
C.~Song, E.~D. Cristofaro, L.~Melis, and V.~Shmatikov, ``Exploiting unintended feature leakage in collaborative learning,'' 2018.

\bibitem{2017Deep}
B.~Hitaj, G.~Ateniese, and F.~Perez-Cruz, ``Deep models under the gan: Information leakage from collaborative deep learning,'' \emph{ACM}, 2017.

\bibitem{2004Efficient}
M.~J. Freedman, K.~Nissim, and B.~Pinkas, ``Efficient private matching and set intersection,'' in \emph{Springer Berlin Heidelberg}, 2004.

\bibitem{2011SMOTE}
K.~W. Bowyer, L.~O. Hall, N.~V. Chawla, and W.~P. Kegelmeyer, ``Smote: Synthetic minority over-sampling technique,'' 2011.

\bibitem{gender-voice}
https://www.kaggle.com/datasets/primaryobjects/voicegender.

\bibitem{company-bankruptcy-prediction}
https://www.kaggle.com/datasets/fedesoriano/company-bankruptcy prediction.

\end{thebibliography}
%\begin{thebibliography}{1}
\bibliographystyle{IEEEtran}
%\end{thebibliography}

%\newpage

\vspace{11pt}
\begin{IEEEbiography}[{\includegraphics[width=1in,height=1.25in,clip,keepaspectratio]{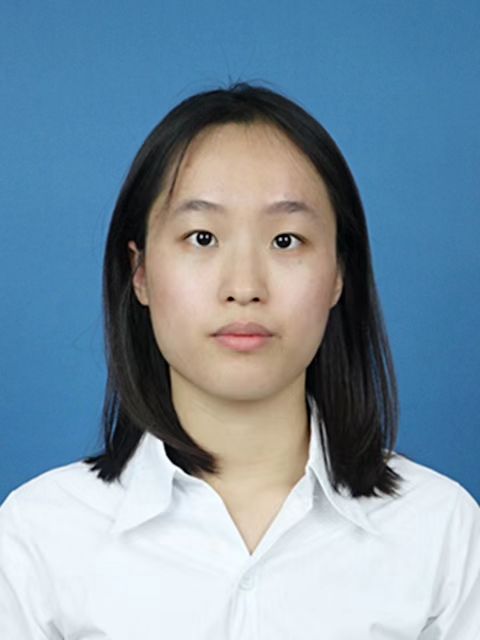}}]{Yuqi Guo}
is currently working toward the MS degree with the School of Computer Information and Technology, Beijing Jiaotong University. Her research interest covers privacy computing.
\end{IEEEbiography}

\begin{IEEEbiography}[{\includegraphics[width=1in,height=1.25in,clip,keepaspectratio]{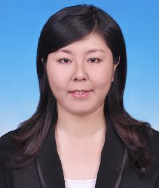}}]{Lin Li}
is currently a professor with the School of Computer and Infor mation Technology, Beijing Jiaotong University. Her current researchinterests include cryptography, privacy computing, security evaluation.
\end{IEEEbiography}

\begin{IEEEbiography}[{\includegraphics[width=1in,height=1.25in,clip,keepaspectratio]{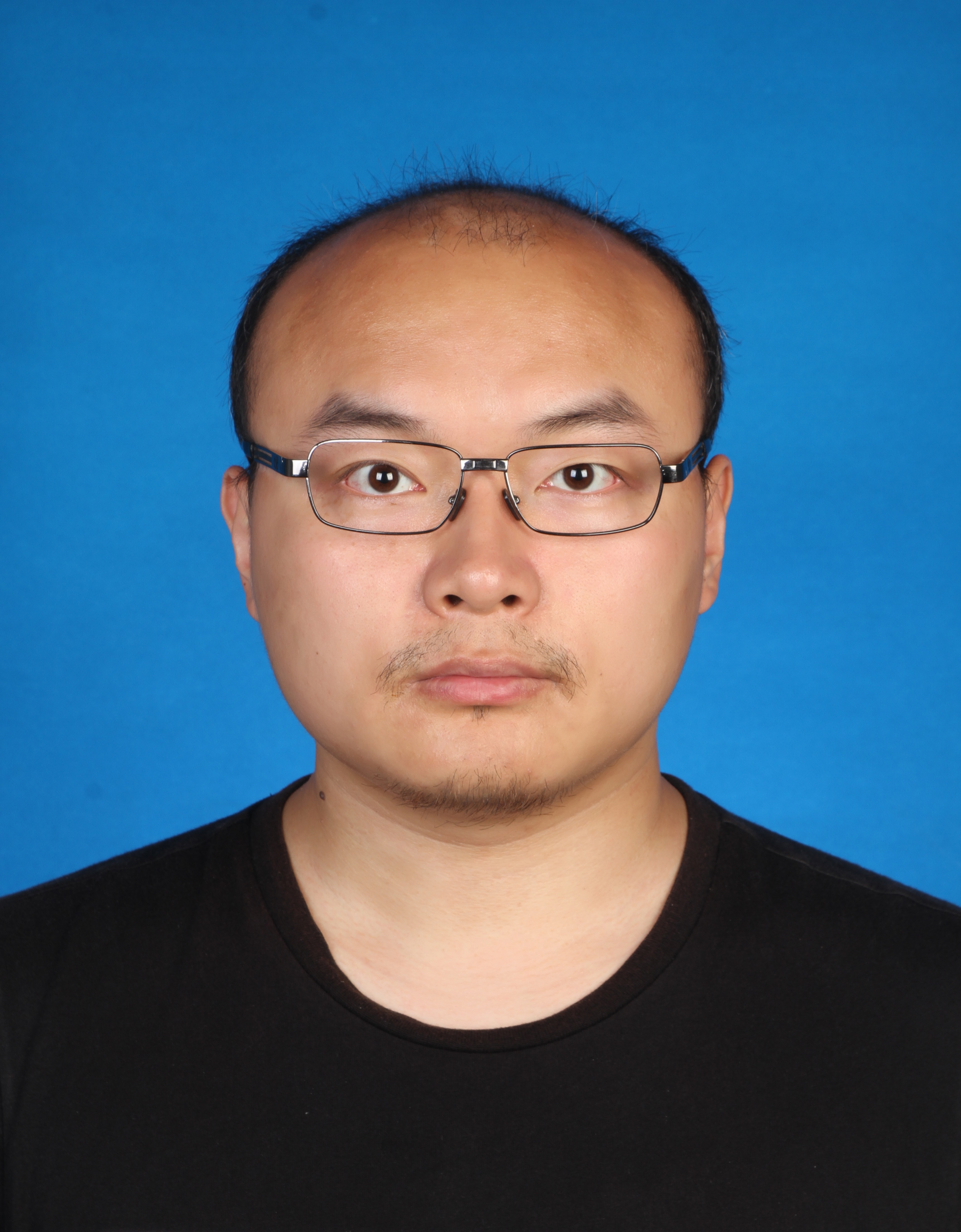}}]{Zhongxiang Zheng}
received his PhD from the Department of Computer Science and Technology, Tsinghua University in 2019 and is currently a lecturer at Communication University of China. He has been engaged in cryptography-related research for a long time, and his main research interests are cryptographic algorithm design and analysis against quantum computing attacks, fully homomorphic encryption algorithm design and analysis, privacy computing, etc.
\end{IEEEbiography}

\begin{IEEEbiography}[{\includegraphics[width=1in,height=1.25in,clip,keepaspectratio]{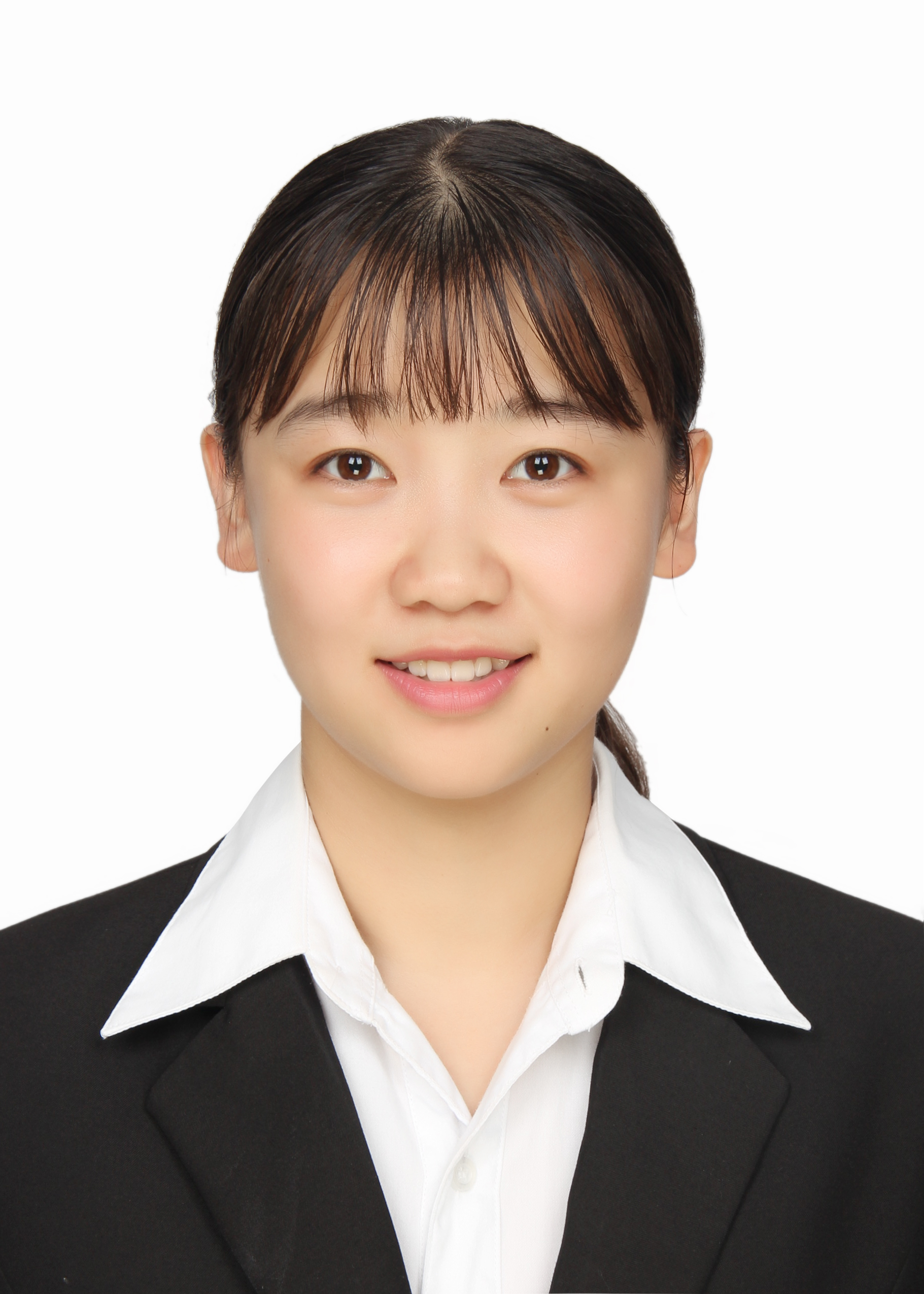}}]{Hanrui Yun}
is currently working toward the MS degree with the School of Computer and Cyber Science, University of Communication University of China. Her research interests include federated learning and homomorphic encryption.
\end{IEEEbiography}

\begin{IEEEbiography}[{\includegraphics[width=1in,height=1.25in,clip,keepaspectratio]{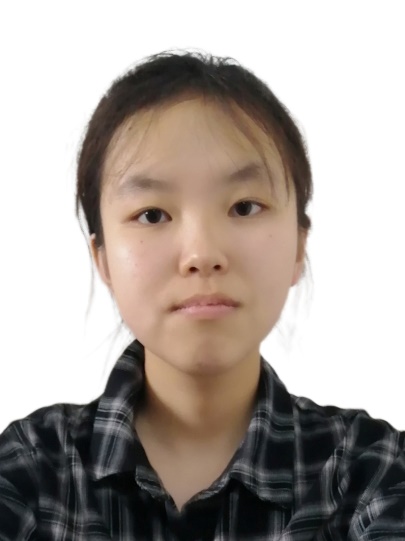}}]{RuoYan Zhang}
is studying for MS degree in the School of Computer and Cyber Sciences, Communication University of China, Beijing, China. Her research interests include the design and analysis of fully homomorphic encryption algorithms, and privacy computing.
\end{IEEEbiography}

\begin{IEEEbiography}[{\includegraphics[width=1in,height=1.25in,clip,keepaspectratio]{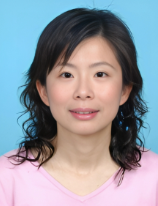}}]{Xiaolin Chang}
(Senior Member, IEEE) is currently a professor with the School of Computer and Infor mation Technology, Beijing Jiaotong University. Her current researchinterests include cloud-edgecom puting, network security, secure and dependable machinelearning.
\end{IEEEbiography}

\begin{IEEEbiography}[{\includegraphics[width=1in,height=1.25in,clip,keepaspectratio]{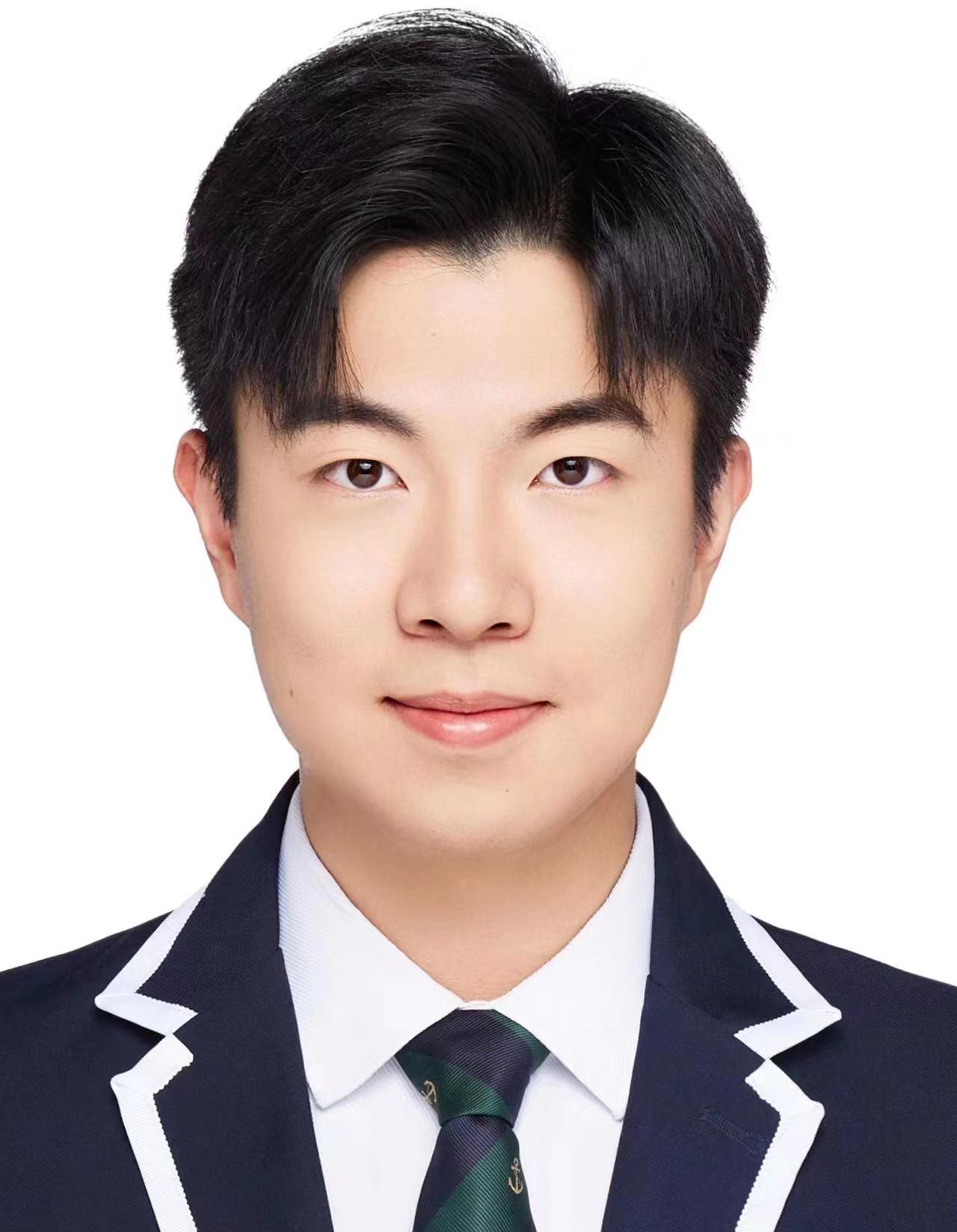}}]{Zhixuan Gao}
is studying for MS degree in the School of Cyberspace Security, Beijing Institute of Technology, China. His research interests include side-channel analysis and network security.
\end{IEEEbiography}
\vspace{11pt}


\vfill
\end{document}